\documentclass[%
 reprint,
nofootinbib,
 amsmath,amssymb,
 aps,
]{revtex4-2}

\usepackage{graphicx}
\usepackage{dcolumn}
\usepackage{bm}

\begin{document}

\preprint{APS/123-QED}

\title{The Effect of Tornadic Supercell Thunderstorms on the Atmospheric Muon Flux}

\author{William Luszczak}
\affiliation{%
 Dept. of Astronomy, Ohio State University\\
}%
\affiliation{
Dept. of Physics and Center for Cosmology and Astro-Particle Physics, Ohio State University
}%

\author{Leigh Orf}
\affiliation{
Cooperative Institute for Meteorological Satellite Studies, University of Wisconsin-Madison
}%

\date{\today}

\begin{abstract}
    Tornadoes are severe weather phenomena characterized by a violently rotating column of air connecting the ground to a parent storm. Within the United States, hundreds of tornadoes occur every year. Despite this, the dynamics of tornado formation and propagation are not particularly well understood, in part due to the challenge of instrumentation: many existing instruments for measuring atmospheric properties are in-situ detectors, making deployment in or near an active or developing tornado difficult. Here, we combine local atmospheric and cosmic ray air shower simulation to explore the potential for remote measurement of the density field within tornado-producing supercell thunderstorms by examining directional variations of the atmospheric muon flux. 
\end{abstract}

\maketitle


\section{Introduction}
\subsection{Tornadoes}

Tornadoes are violently rotating columns of air connecting a cumuliform cloud with the ground. Though tornadoes are relatively common in the United States, the process by which they form is currently not well understood, though significant efforts in theoretical and computational modeling as well as field observations have advanced our understanding a great deal. It is known that the strongest, longest-lived tornadoes form from supercell thunderstorms, defined by the presence of a persistent, rotating updraft. Ground level rotation is intensified by an upward-directed pressure gradient force, converging and amplifying vertical vorticity and producing a region of intense circulation~\cite{10.1063/PT.3.2514}. Tornadoes are typically embedded within a region of larger scale circulation referred to as a \textit{mesocyclone}. While the mesocyclone encompasses a large volume of lowered air density and atmospheric pressure, the most extreme pressure and density perturbations are found within the tornado itself~\cite{mesocyclone}. 

Recent simulations of tornado formation suggest the presence of a low pressure core in the parent supercell thunderstorm that contributes to tornado formation~\cite{EvolutionSimulation}, however experimental measurements of this region are logistically difficult. Current pressure field measurements rely on deploying instrumentation in the path of an approaching storm~\cite{InSituMeasurements}. This method can be unreliable, as it requires a tornado to track near the pre-deployed sensors, and strong tornadoes may damage or destroy the instrument. Surface-based measurements are also only able to measure the pressure field at ground level, concealing variations in the pressure field with altitude. Drone and aircraft based measurements can measure the pressure field as at function of altitude~\cite{torus}, but these measurements are restricted to the immediate vicinity of the instrument itself, making measurements over large volumes and within turbulent weather systems difficult. 

\subsection{Atmospheric Muons}
Cosmic rays are charged particles (protons, as well as heavier atomic nuclei) accelerated in faraway astrophysical sources. Though the exact origins of cosmic rays remain an open question in astrophysics, their flux at Earth is well measured~\cite{Workman:2022ynf}. Cosmic rays that reach Earth can interact with nuclei in the atmosphere, producing a shower of secondary particles. This particle shower includes pions and kaons, which will eventually decay, producing muons:

\begin{equation}
\pi^{\pm} \rightarrow \mu^{\pm} + \overset{\textbf{\fontsize{2pt}{2pt}\selectfont(---)}}{\nu}_{\mu}
\end{equation}
\begin{equation}
K^{\pm} \rightarrow \mu^{\pm} + \overset{\textbf{\fontsize{2pt}{2pt}\selectfont(---)}}{\nu}_{\mu}
\end{equation}

Muons also lose energy as they propagate through matter. Their energy losses can be described by:

\begin{equation}
\frac{dE}{dx}=\rho(x)[a(x,E)+ E*b(x,E)]
\end{equation}

\noindent where E is the energy of the muon, $x$ is the distance traveled, $\rho(x)$ is the density of the traversed matter and $a(x, E)$, $b(x, E)$ are the ionisation and radiative energy losses. The quantities $\rho$, $a$, and $b$ are all dependent on the properties of the matter traversed. For atmospheric muons, $\rho$ is directly proportional to the atmospheric pressure~\cite{LECHMANN2021103842}, suggesting that muographic measurements of air density could be used to infer the atmospheric pressure in a region. 

Muons eventually decay, producing electrons and neutrinos:

\begin{equation}
\mu^{-} \rightarrow e^{-} + \bar{\nu}_{e} + \nu_{\mu}
\end{equation}
\begin{equation}
\mu^{+} \rightarrow e^{+} + \nu_{e} + \bar{\nu}_{\mu}
\end{equation}

Muons have a lifetime of approximately 2.2 $\mu s$ in the rest frame~\cite{Workman:2022ynf}. In the frame of the surface of the Earth, energetic muons are time dilated, allowing them to travel significant distance ($\approx $10s to 100s of kilometers) into the atmosphere before decaying. Higher energy muons are able to travel farther prior to decaying, and in combination with the energy losses due to matter described above, this means that the overall atmospheric muon flux is effectively attenuated by matter.

In the past, this effect has been leveraged as a probe of matter density along the muon path.  A muon detector is placed near an object of interest, and the flux of muons travelling through the object is measured and compared to a nominal flux of muons that did not pass through the object. The suppression of the muon flux observed travelling through the object is directly related to the integral of the density along the muon path. This technique (``muon tomography”) has previously been used in a variety of contexts, including measurements of the interiors of volcanoes and the Great Pyramids~\cite{LECHMANN2021103842}.

The atmospheric muon flux is known to be proportional to the local atmospheric pressure (and thus, atmospheric density as well). This has been demonstrated to be true on large scales: daily and weekly variations in the atmospheric muon flux that are correlated with atmospheric pressure have been observed by various muon detectors~\cite{Jourde_2016,tilav2019seasonal}. On smaller scales, local variations in atmospheric pressure have been observed in data from muon telescopes pointed at typhoons off the coast of Japan~\cite{typhoons}, as well as non-tornadic thunderstorms~\cite{muthunderstorms,muthunderstorms2}.

Previous studies have also identified variations in the atmospheric muon flux associated with electric fields inside thunderstorms~\cite{TASDLightning,muthunderstorms}. For the purposes of this paper, electric field effects are neglected, as we seek to explore solely the effect of the pressure field on the muon flux. In practice, however, both the pressure and electric field would need to be modeled to fully describe any potential real-world observations.

\section{Remote Sensing of Pressure Using Atmospheric Muons}
\subsection{Conceptual Overview}
Since the atmospheric muon flux is correlated with the density of matter along the muon path, and the density of air in the atmosphere is correlated with the atmospheric air pressure, this implies that variations in atmospheric air pressure can be observed as variations in the atmospheric muon flux. A hypothetical muon detector could detect variations in air pressure as a function of altitude by measuring the muon flux at different zenith angles. As mentioned above, this has been done to measure the low pressure region of typhoons, where a region of increased muon flux was found to be correlated with the low-pressure regions of the target typhoon~\cite{typhoons}. Given the success of this method in the context of typhoon measurements, we consider extending this technique to make a similar measurement of the density field within supercell thunderstorms. 

In comparison to typhoons, supercell thunderstorms present several unique challenges. Supercell thunderstorms are significantly smaller than typhoons. While typhoons can grow to be hundreds of kilometers in diameter, supercell thunderstorms typically only occupy an area tens of kilometers in radius. The smaller physical scale of supercell thunderstorms translates to a smaller effect on the muon flux, as muons are in a low-density region for a smaller portion of their total propagation path. This manifests as a smaller difference in atmospheric muon flux when comparing the high and low pressure regions of the storm. Additionally, supercell thunderstorms evolve quickly, with lifetimes rarely exceeding 4 hours~\cite{supercellevolution}. The short window of observation needed to observe the dynamics of these storms makes it difficult to collect enough muon data to observe small variations in the muon flux. This effect can be mitigated by simply constructing a larger muon detector, as this allows for a larger number of muon counts to be observed over the same time period. Thus, there are two important questions that we seek to answer:

\begin{itemize}
  \item{Given an estimate of the density perturbations within a supercell thunderstorm, how large is the corresponding effect on the atmospheric muon flux?}
  \item{How large of a muon detector is needed to observe an effect of this scale on reasonable ($\approx$hour) time scales?}
\end{itemize}

For the purposes of this study, a setup similar to that described in reference~\cite{typhoons} is simulated. A muon detector is simulated on the ground 10 kilometers from the center of a thunderstorm. Cosmic ray showers are generated in the upper atmosphere, and propagated through the thunderstorm to the location of our muon detector. The storm is examined at 6 different times, summarized in table \ref{tab:sim_table}. No forward motion of the storm is simulated, so the relative position of the storm relative to the muon detector is fixed. The observed number of muon counts over 1 hour is recorded along various lines of sight leading to the detector, producing a ``map" of muon counts as a function of elevation ($\theta$) and azimuthal ($\phi$) angle. A similar map is assembled using the same setup without the presence of a thunderstorm, and the two maps are compared to determine the effect of the storm on the atmospheric muon flux. 

\begin{center}
\begin{table}
\begin{tabular}{||c c||} 
 \hline
 \textbf{Time [s]} & \textbf{Storm State} \\ [0.5ex] 
 \hline\hline
 3600 & No tornado  \\ 
 \hline
 5100 & No tornado  \\
 \hline
 5580 & Tornado recently formed  \\
 \hline
 7200 & Strong tornado \\
 \hline
 8040 & Strong tornado \\
 \hline
 10200 & Tornado has dissipated \\ [1ex] 
 \hline
\end{tabular}
\caption{A summary of the storm simulation at the 6 snapshots in time that were investigated. The number of muon counts observed at each of the listed times is calculated by integrating the muon flux over the preceding hour.}
\label{tab:sim_table} 
\end{table}
\end{center}

\begin{figure*}
\includegraphics[width=0.5\textwidth]{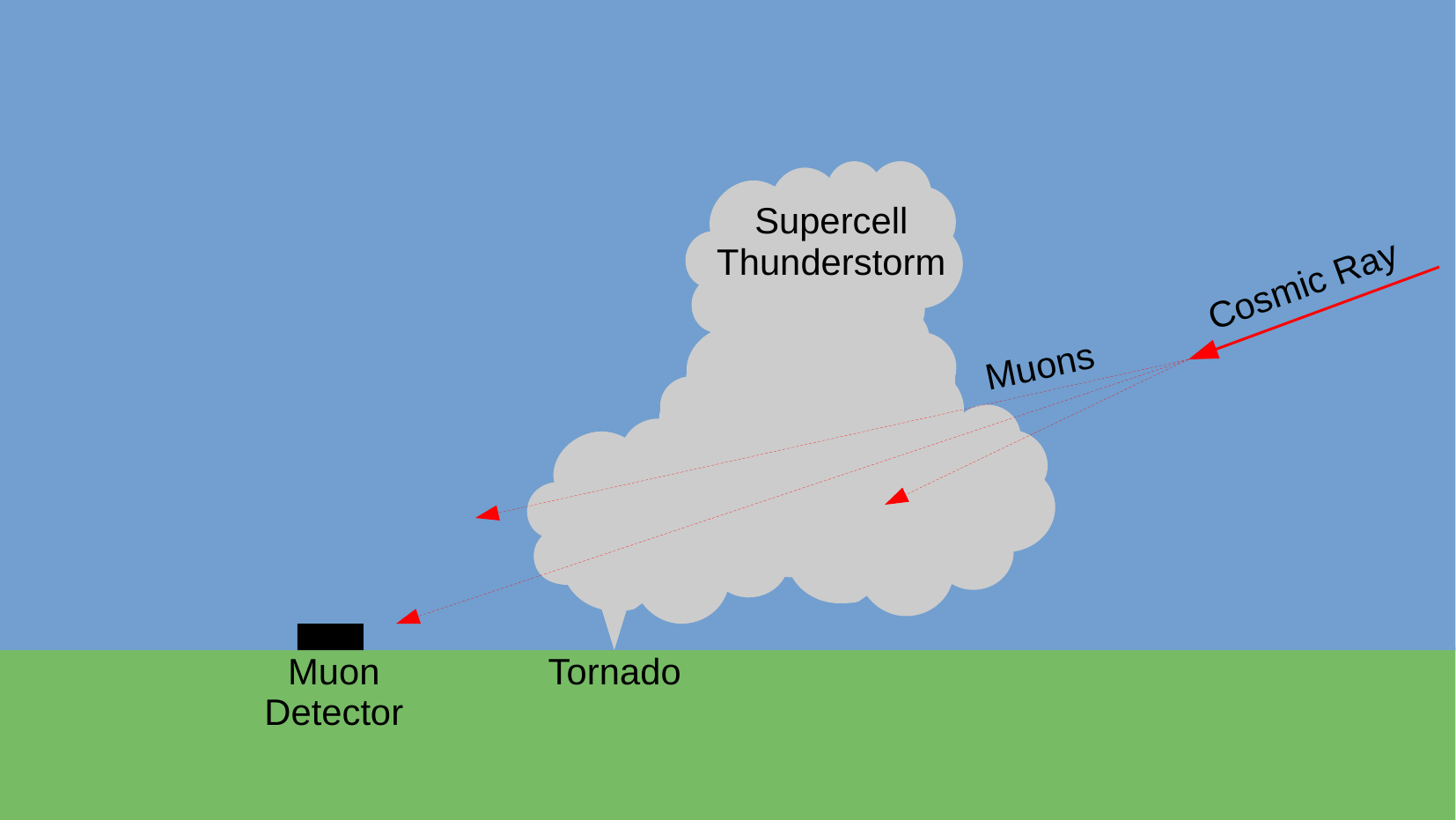}
\includegraphics[width=0.5\textwidth]{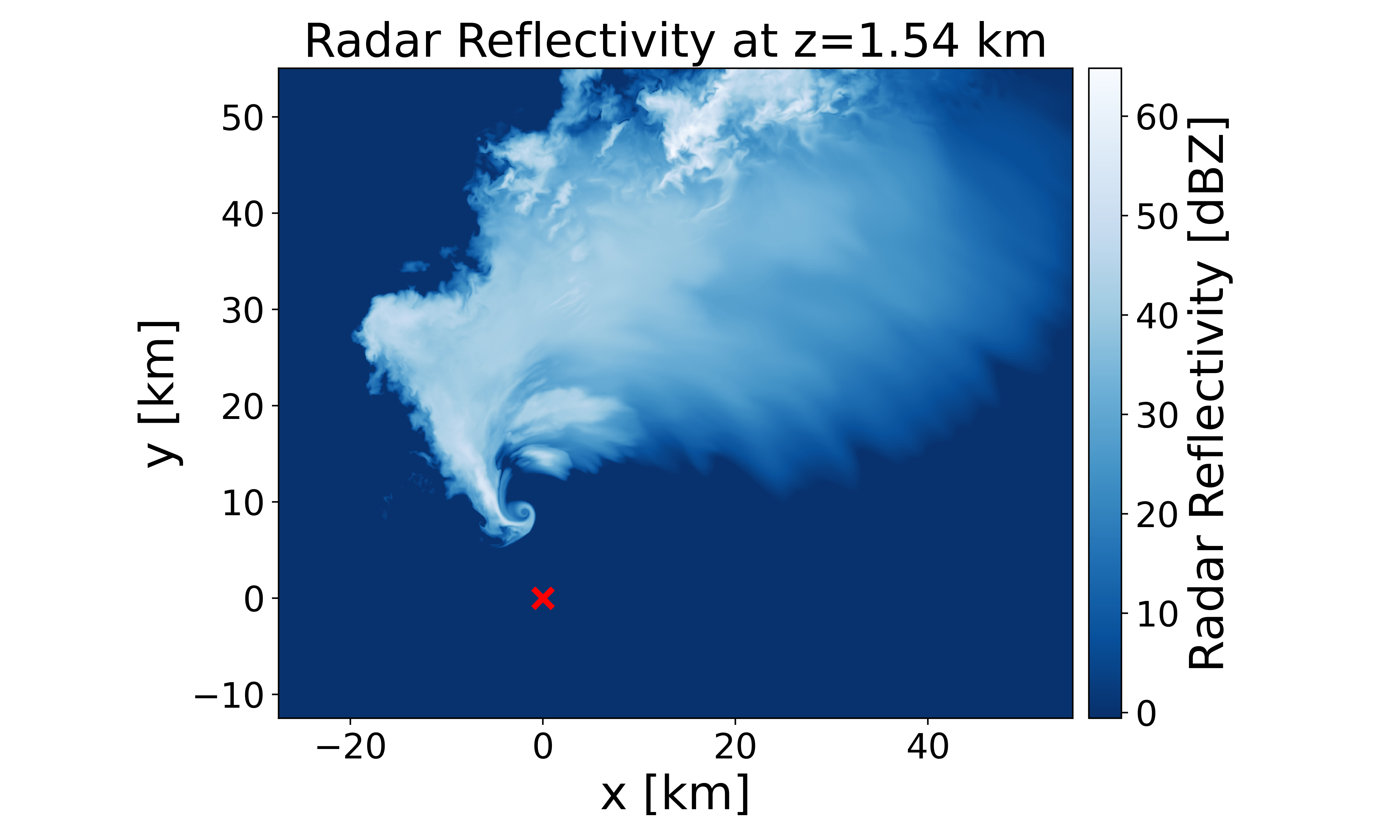}
\caption{\label{fig:setup_sketch} Left: A cartoon of the setup considered in this paper (not to scale). Cosmic rays interact in the upper atmosphere, and secondary muons propagate through a supercell thunderstorm. A muon detector is placed on the opposite side of the storm, and the effect of the thunderstorm  density field on the flux of atmospheric muons is calculated. Right: A top-down view of the simulated arrangement, showing the location of the muon detector as a red ``x", as well as the simulated storm radar reflectivity at an altitude of 1.54 km. The tornado is located in the hook echo near $(x,y)=(0,10)$}
\end{figure*}

\subsection{Supercell Thunderstorm Simulation} \label{stormsim}

\begin{figure}
    \centering
    \includegraphics[width=0.45\textwidth]{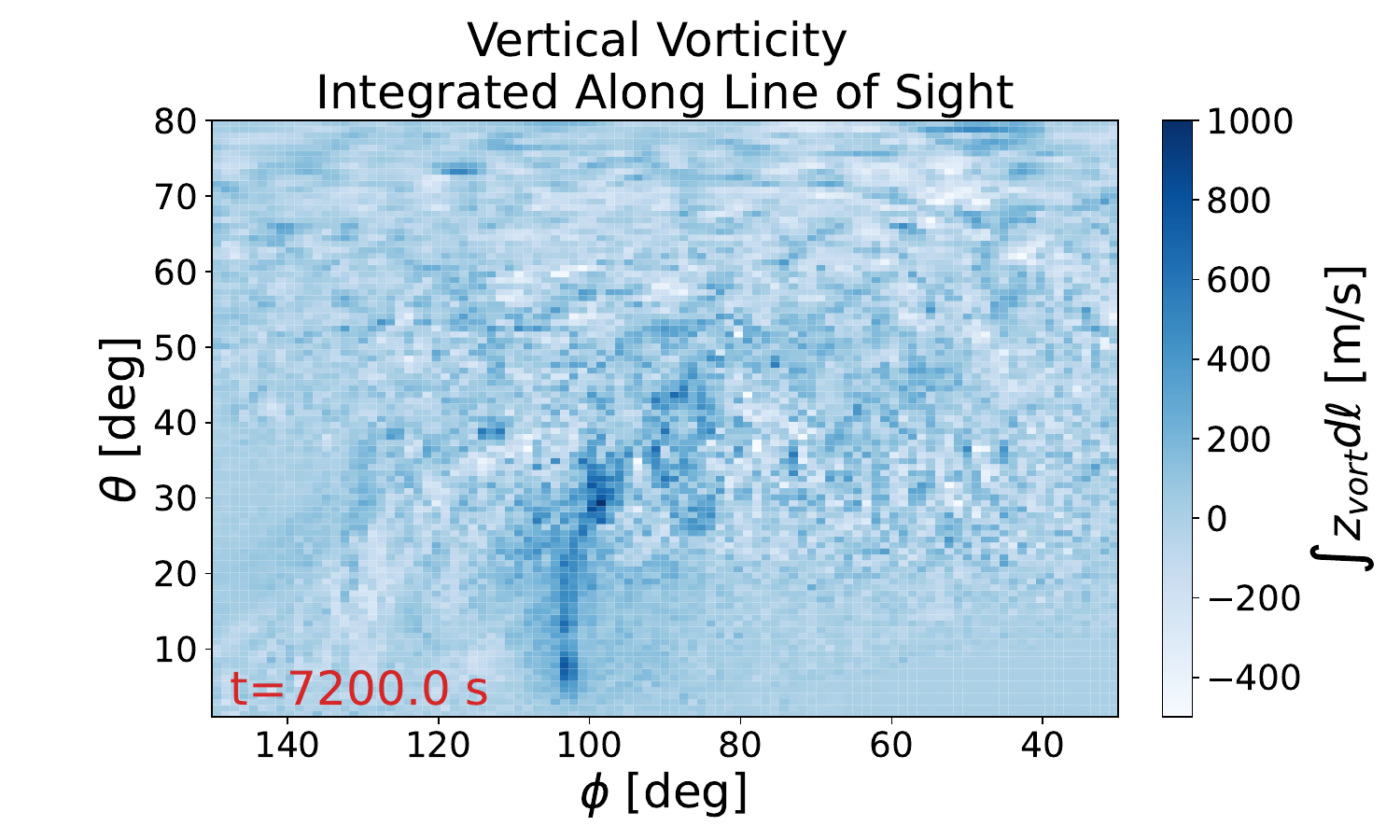}
    \caption{The vertical component of vorticity integrated along lines of sight leading to the muon detector at $t=7200$ seconds. The tornado can be seen as the feature producing high vertical vorticity near $\phi=100$ degrees.}
    \label{fig:zvort}
\end{figure}

Cloud Model 1 (CM1), a three-dimensional nonhydrostatic fully compressible cloud
model, was used to simulate the tornadic supercell
thunderstorm~\cite{Bryan2002-ie}. Among the prognostic variables CM1 solves for
include wind, potential temperature, pressure, turbulent kinetic energy, dry air density, total density and
water substance including water vapor, cloud water, cloud ice, rain, snow,
graupel and hail. For the purpose of this study, only the total density field was
analyzed. The initial environmental conditions for the simulation were identical
to \cite{Orf2017-eu} and \cite{Orf2019-kn}, namely, the state of the atmosphere
prior to the observed 24 May 2011, El Reno, Oklahoma supercell that produced a
long-track violent tornado. Physics parameterization options for the simulation
include three-moment microphysics \cite{Mansell2010-uq} and a subgrid turbulent
kinetic energy closure scheme \cite{Deardorff1980-lx}. The simulation utilized
an isotropic cubic mesh with a grid spacing of 75\,m, and was triggered using an
updraft nudging technique \cite{Naylor2012-ml}. The simulation morphology is
similar to \cite{Orf2017-eu}, with the storm initially splitting with a dominant
right-moving supercell (analyzed herein) producing a vigorous low-level rotating
updraft (mesocyclone) prior to the formation of a violent tornado. The density
field of the storm is analyzed at six model times, detailed in table \ref{tab:sim_table}. A tornado forms at $t=5580$ seconds, producing a region of low density (figure \ref{fig:integrateddensity}), and a region of high vertical vorticity (seen in figure \ref{fig:zvort}).

\subsection{Muon Simulation}
MCEq\footnote{\url{https://github.com/mceq-project/MCEq/tree/master/MCEq}} is used to numerically solve the atmospheric particle cascade equations describing the muon flux as the shower propagates through the atmosphere. The primary cosmic ray flux is simulated according to the model described in \cite{h3acitation}. Primary cosmic ray interactions are then simulated using the SIBYLL 2.3c interaction model~\cite{sibyll}. 

A simulated muon detector is placed at the origin, with local atmospheric variations obtained from the model described in the previous section simulated in a 37.5 km (length) $\times$ 37.5 km (width) $\times$ 28.05 km (height) box surrounding the detector. Outside this region we use a standard unperturbed atmospheric model identical to the US standard atmosphere implemented in CORSIKA\footnote{\url{https://www.iap.kit.edu/corsika/98.php}}, originally parameterized by J. Linsley~\cite{corsika}. 

The combined number of $\mu^{+}$ and $\mu^{-}$ counts for each $2 \times 2$ degree angular bin in $\theta$ and $\phi$ is calculated by integrating the muon flux. The calculations presented here assume a muon detector with perfect efficiency ($\epsilon=1.0$). Modern muon detectors have been shown to have efficiencies of at least $\epsilon=0.5$~\cite{deteff,muondeteff_slac,ANGHEL201512}. Smaller muon detection efficiencies would introduce a factor of $\epsilon$ to equations \ref{eq:nexp} and \ref{eq:nobs}. 

Using $\Phi_0$ to denote the muon flux under an unperturbed (no-storm) atmosphere, the nominal expected number of events would then be:

\small
\begin{equation}
   N_{exp}(\theta_{bin}, \phi_{bin}) =  \int_{\Delta t} \int_{E_{min}}^{E_{max}} \int_{\Delta\phi}\int_{\Delta\theta}  \Phi_{0}(\theta, \phi, E) \,d\Omega dE dt
\label{eq:nexp}
\end{equation}

\noindent where $\Delta \theta$ and $\Delta \phi$ correspond to the size of the angular bins in azimuth and elevation (2 degrees). The bounds of the energy integral, $E_{min}$ and $E_{max}$ describe the minimum and maximum muon energies that were simulated. For this paper, muon fluxes were calculated between $E_{min}=0 \text{ TeV}$ and $E_{max}=10^8 \text{ TeV}$, however the vast majority ($>99\%$) of the atmospheric muon flux reaching the ground exists between 0 and 1 TeV. 

We choose a total integration time of $\Delta t$ = 1 hour, examining the storm at set times using simulated muon data from the preceding hour. The muon flux is integrated in tandem with the simulated storm evolution, with temporal bin edges defined by the times listed in table \ref{tab:sim_table}. The density field is considered to be fixed within each temporal bin, though all bins are smaller than the total integration interval of 1 hour.

To give a sense of scale: for a 1000 square meter muon detector, the expected rate of atmospheric muons is approximately 28,400 events per square degree angular bin per hour at $\theta=40^{\circ}$. This rate varies naturally as a function of elevation angle, with a lower flux near the horizon and a higher flux from very vertical elevation angles~\cite{GaisserCR}. 

An expectation of the observed number of counts can be calculated by integrating the muon flux resulting from simulating the atmospheric model described in the previous section ($\Phi_{storm}$): 

\small
\begin{equation}\label{eq:nobs}
\begin{split}
N_{obs}(\theta_{bin}, \phi_{bin}) = {}\\
\int_{\Delta t} \int_{E_{min}}^{E_{max}} \int_{\Delta\phi}\int_{\Delta\theta} \Phi_{storm}(\theta, \phi, E, t) \,d\Omega dE dt
\end{split}
\end{equation}

Poisson variations are added to $N_{obs}$ to simulate real data. Since the muon rate varies by several orders of magnitude as a function of elevation angle, it is convenient to work with a Poisson likelihood describing the probability of observing $N_{obs}$ counts from a particular direction:

\begin{equation}
  L(N_{obs}) = \int_{N_{obs}}^{\infty} P(N_{exp}, x) dx
\end{equation}

where $P(N_{exp}, x)$ is a Poisson distribution with mean $N_{exp}$, obtained from equation \ref{eq:nexp}. This normalizes the resultant maps of the muon flux to the expected rate under an unperturbed atmosphere at a particular elevation angle so muon paths at different elevation angles are easier to compare. $L$ has been constructed to be proportional to the atmospheric air density. As the air density decreases, the muon flux increases, resulting in a lower value of $L$.

\section{Results}
\begin{figure*}[ht]
    \includegraphics[width=0.4\textwidth]{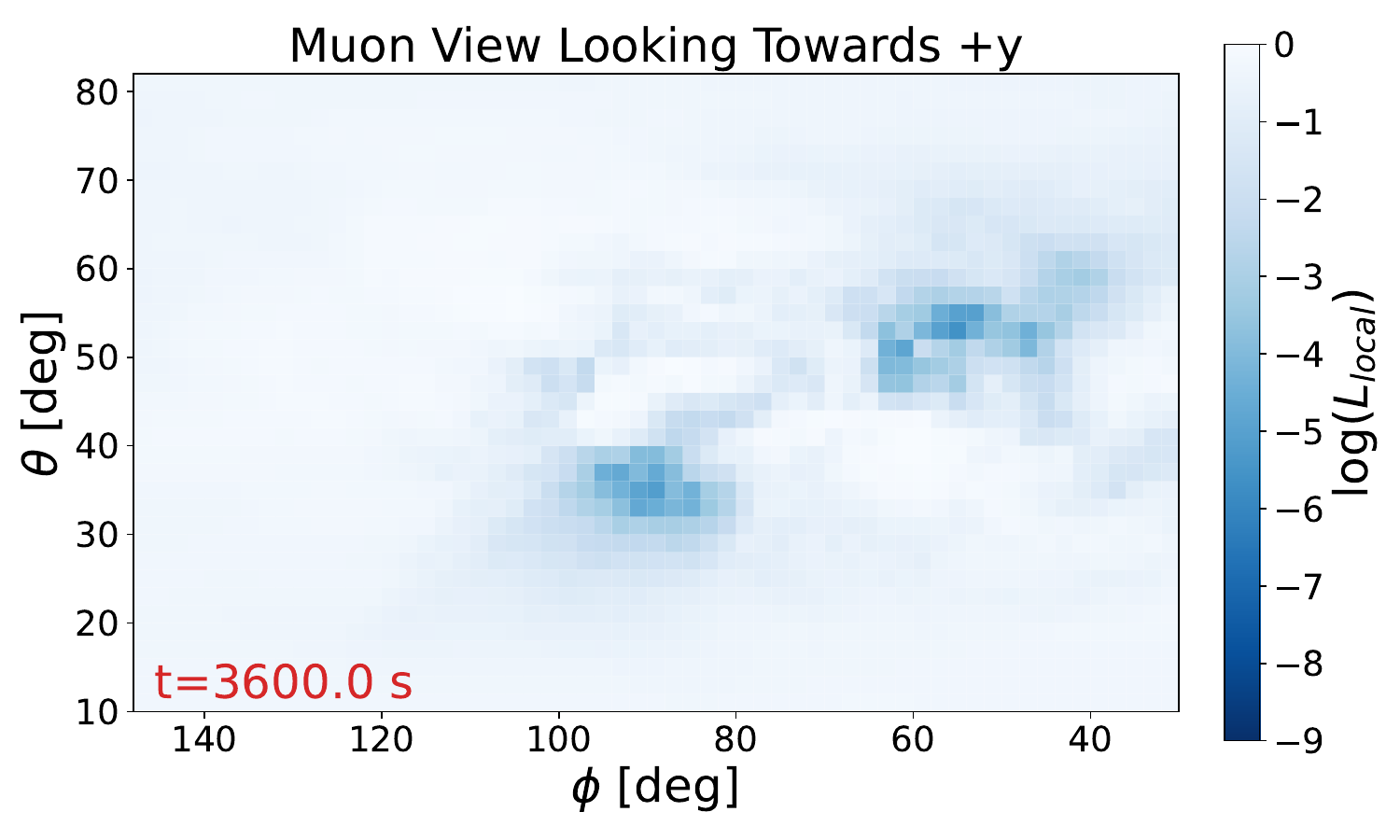}
    \includegraphics[width=0.4\textwidth]{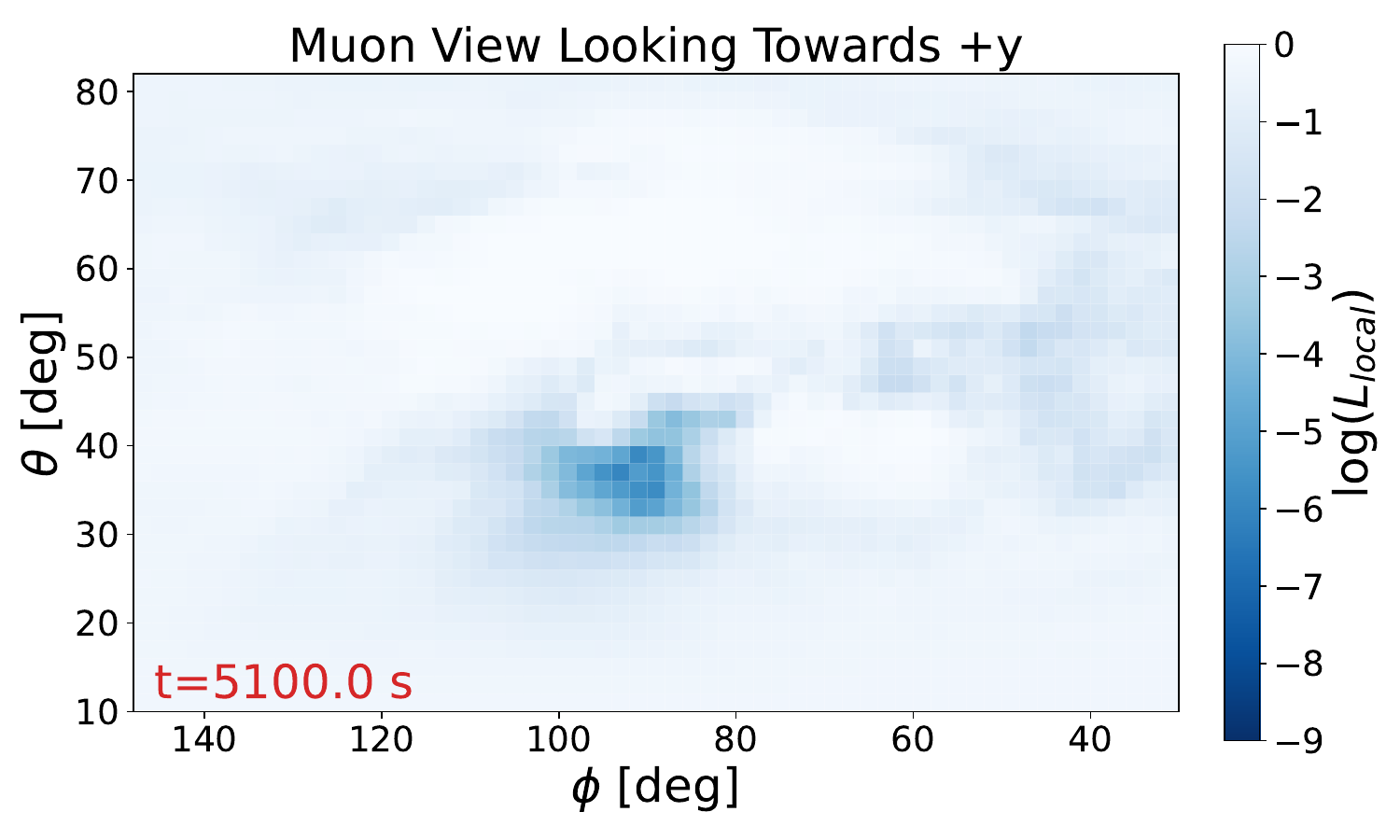}
    \includegraphics[width=0.4\textwidth]{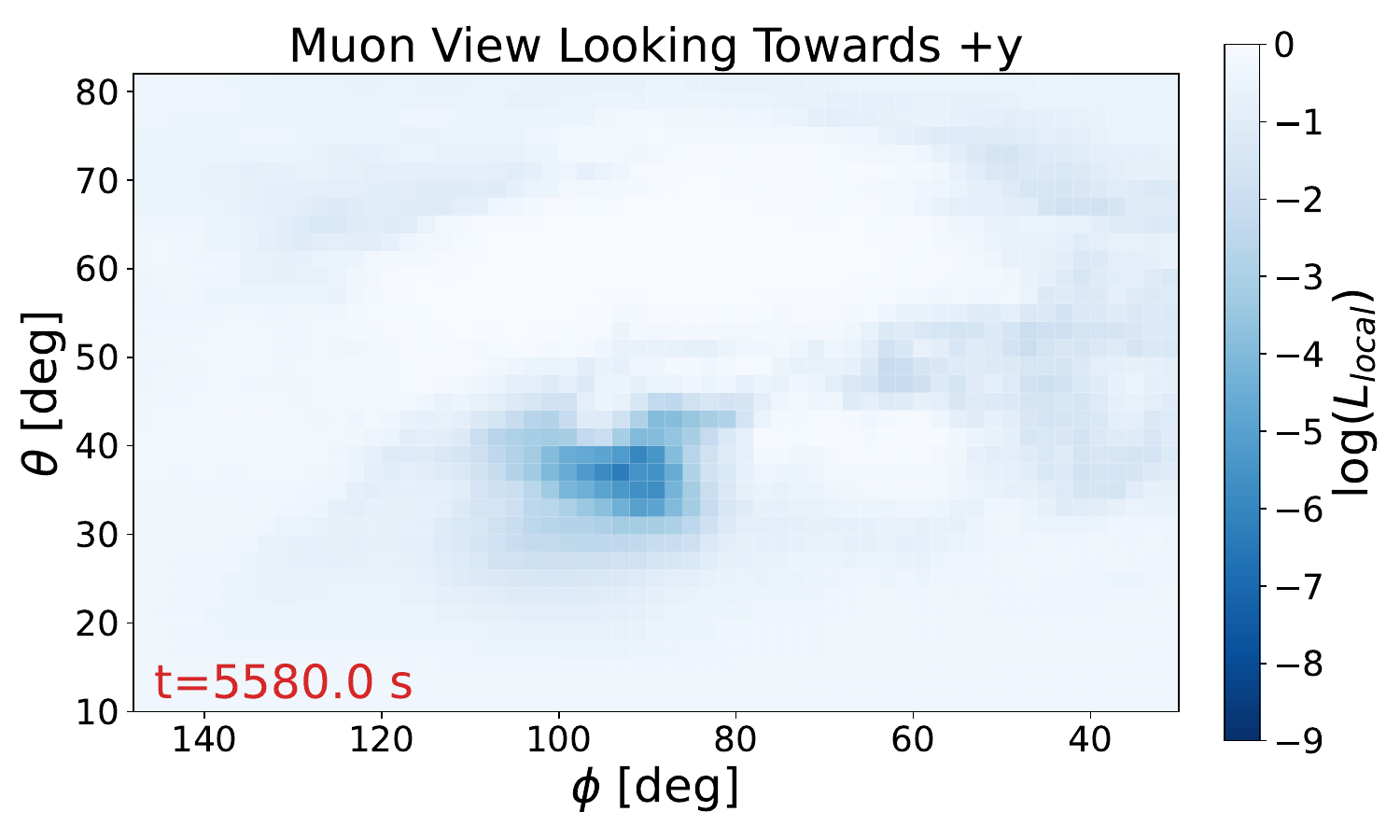}
    \includegraphics[width=0.4\textwidth]{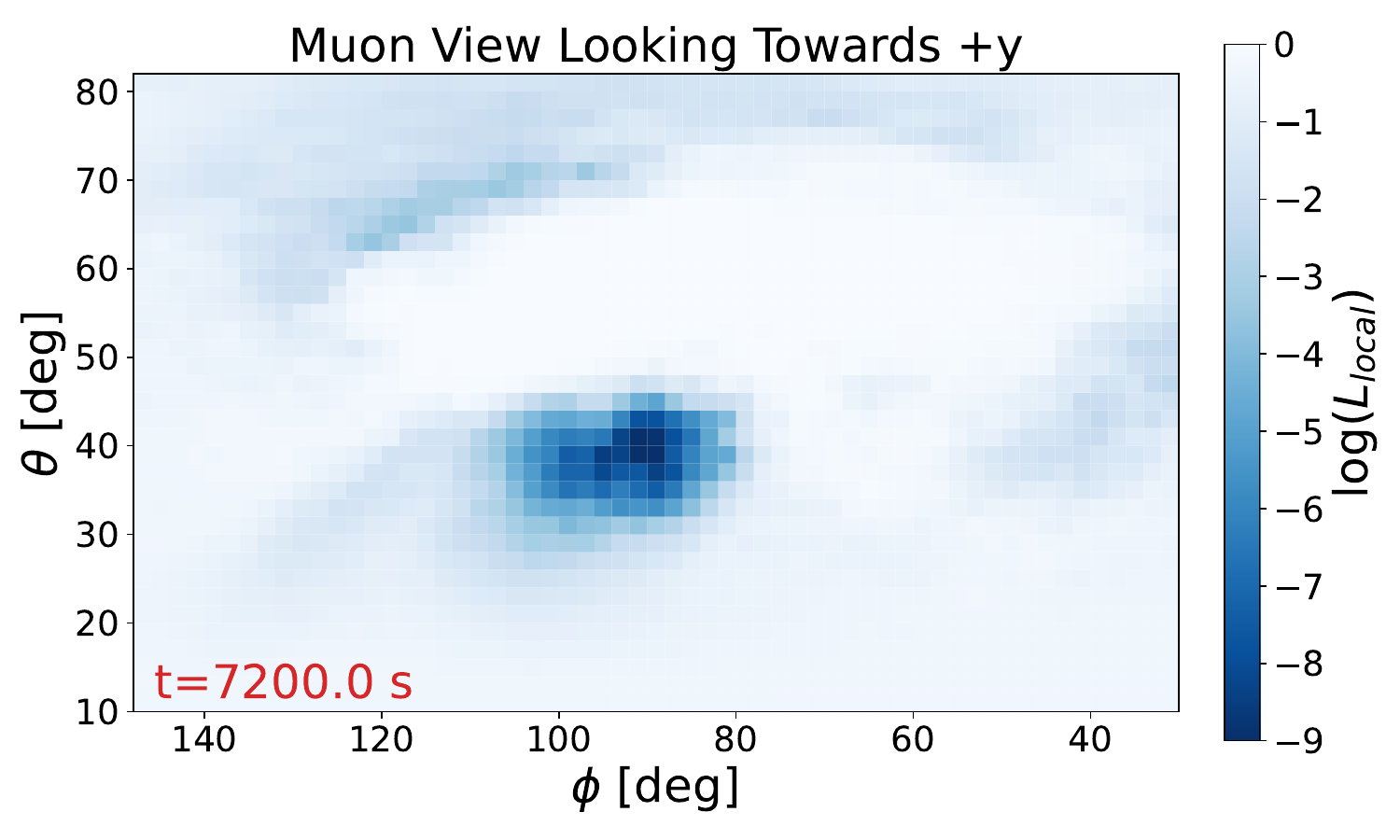}
    \includegraphics[width=0.4\textwidth]{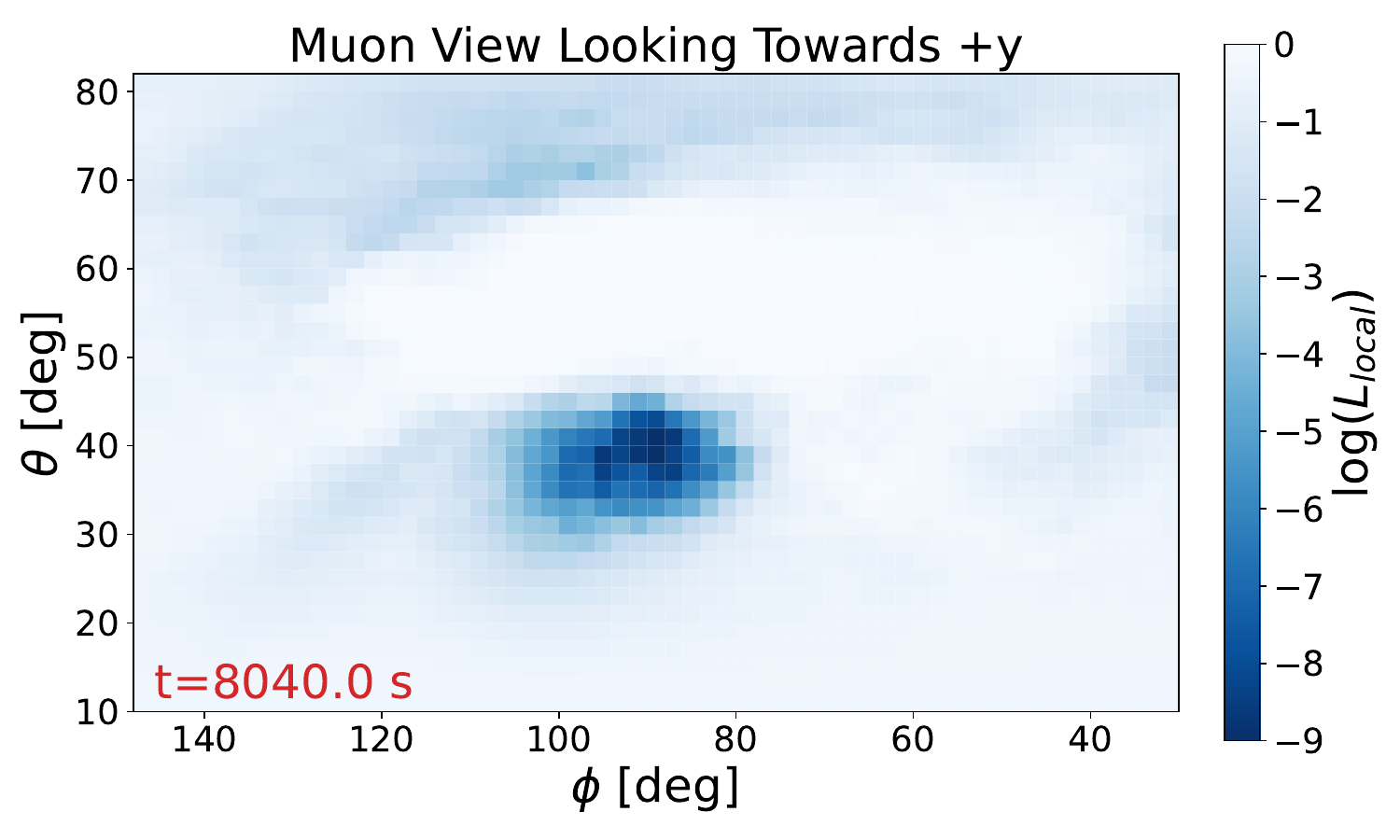}
    \includegraphics[width=0.4\textwidth]{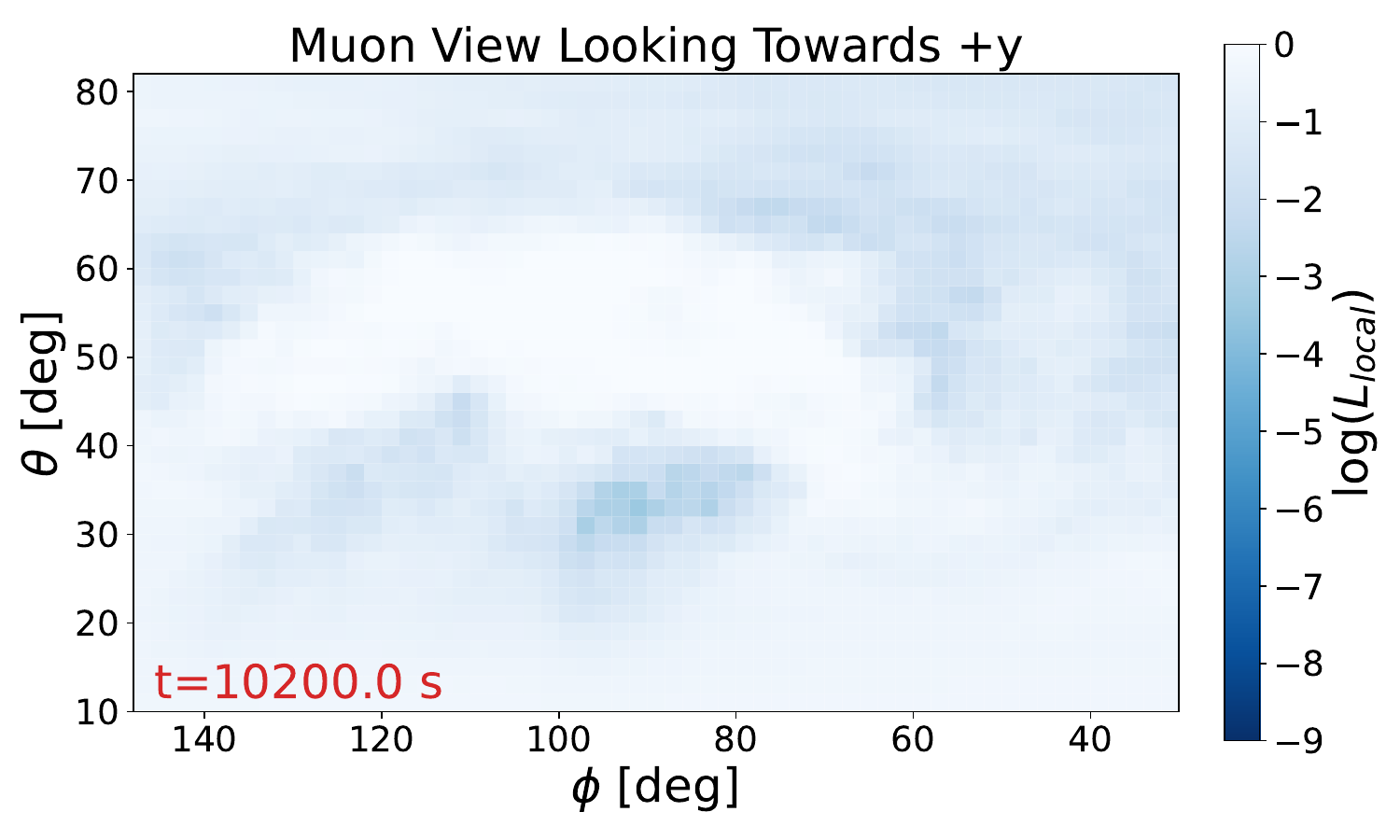}
\caption{\label{fig:muonflux}Idealized (without Poisson noise) expectations of the effect of a supercell thunderstorm on the atmospheric muon flux. Each image integrates the muon flux over the preceding hour. The tornado forms at $t=5580$ seconds, and a corresponding intensification of the muon flux perturbation at $\phi=100$ degrees can be seen following this time.}
\end{figure*}

Figure \ref{fig:muonflux} shows the idealized expectation of the effect of the storm density field on the atmospheric muon flux along various lines of sight for a 1000 m$^2$ muon detector, over the course of 1 hour preceding the time listed for the map. These maps show the observed atmospheric muon flux as a function of azimuthal angle ($\phi$), and elevation angle ($\theta$). $\phi=0^{\circ}$ is defined to point parallel to the positive x-axis, and increases in a counterclockwise direction (so $\phi=90^{\circ}$ is parallel to the positive y-axis).  Differences in $\phi$ and $\theta$ can be converted to approximate horizontal and vertical lengths given a distance to the region of interest using trigonometry:

\begin{equation}
  d_{horizontal} = 2 R \text{tan}(\Delta \phi)
\end{equation}
\begin{equation}
  d_{vertical} = 2 R \text{tan}(\Delta \theta)
\end{equation}

where $d_{horizontal}$ and $d_{vertical}$ are the horizontal and vertical distances between two points, $R$ is the distance between the observer and the region of interest, and $\Delta \phi$ and $\Delta \theta$ are differences in $\phi$ and $\theta$ for two arbitrary points on the maps shown in figures like figure \ref{fig:muonflux}.

Regions of increased muon flux, identifiable as groups of pixels with lower values of $L$, correspond to low-density regions of the storm. Similarly, regions of suppressed muon flux correspond to higher-density regions. Note, however, that the muon flux is correlated with the average of the density field along a particular line of sight, meaning information about the density field as a function of depth is not easily accessible from the two dimensional images shown. Depth-dependent information about the density field could conceivably be obtained by multiple detectors with different views of the same storm, or by examining the energy distribution of detected muons, however development of this idea is left for future study.

\begin{figure*}
    \includegraphics[width=0.4\textwidth]{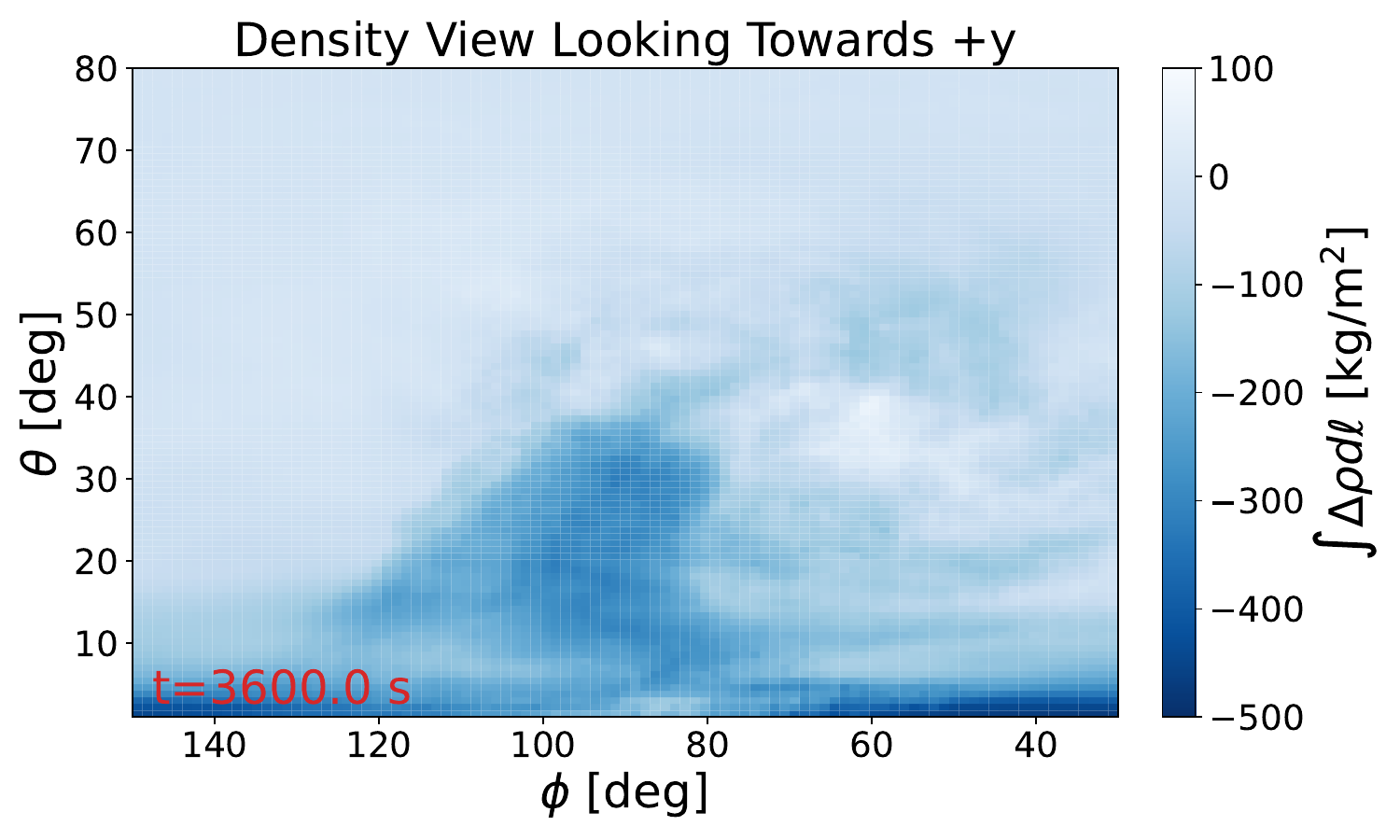}
    \includegraphics[width=0.4\textwidth]{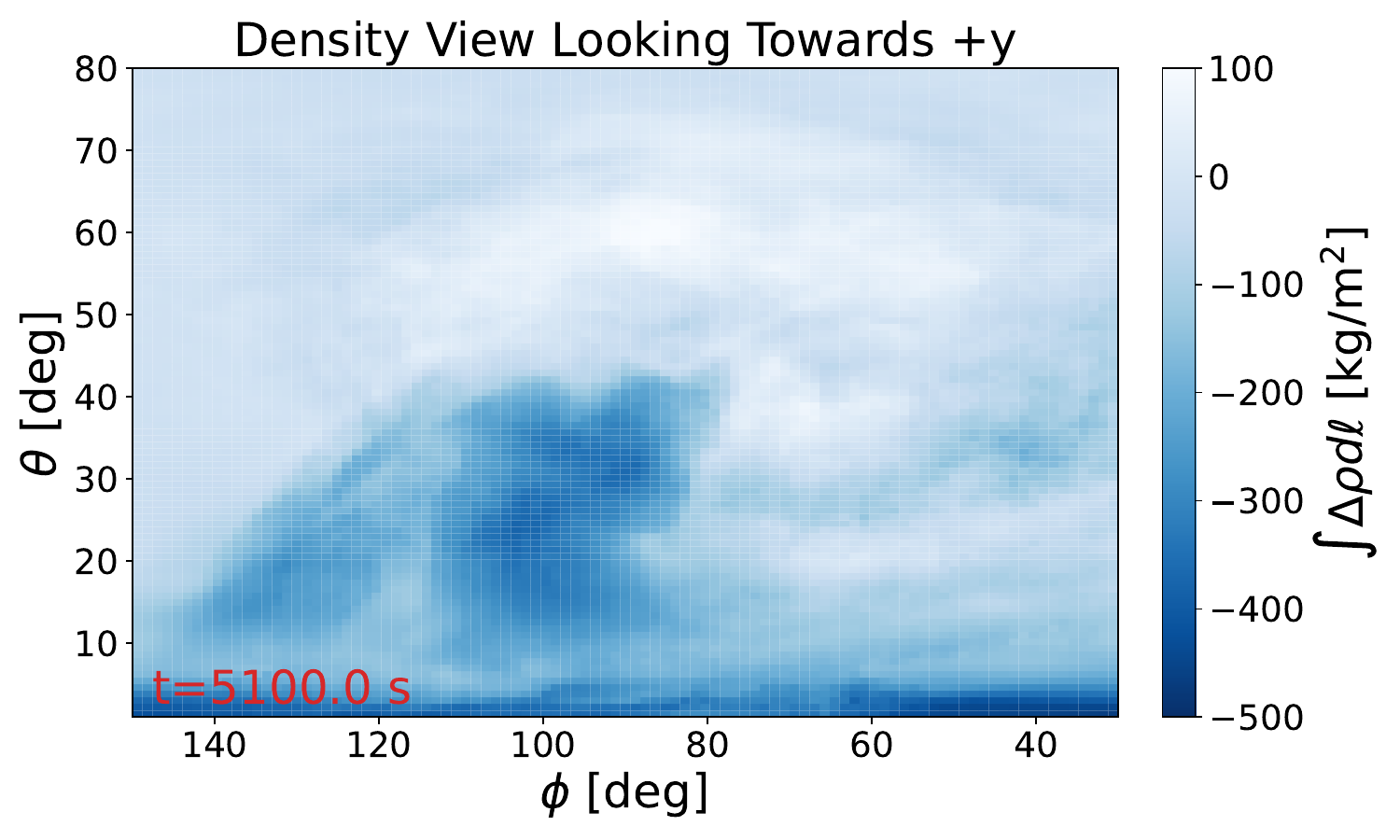}
    \includegraphics[width=0.4\textwidth]{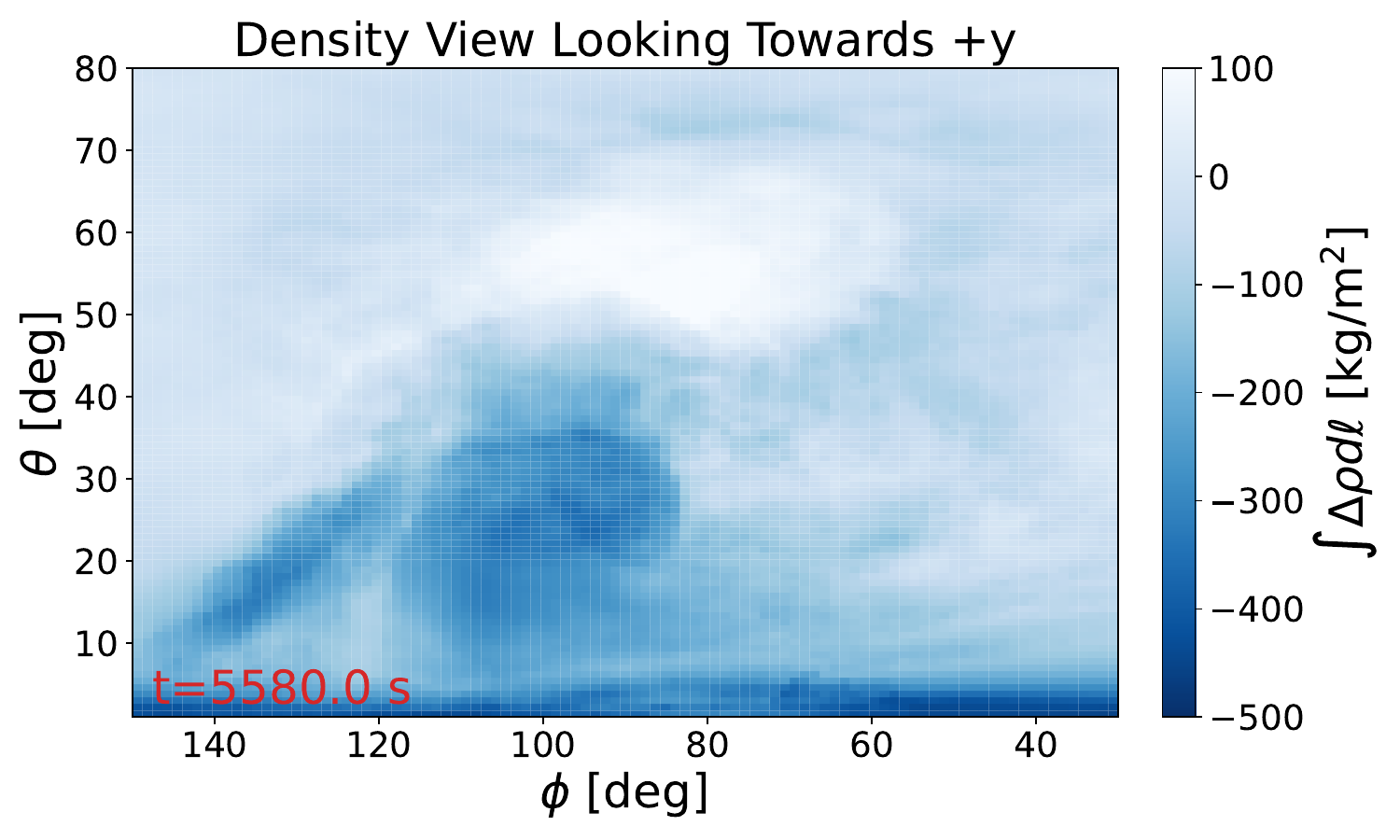}
    \includegraphics[width=0.4\textwidth]{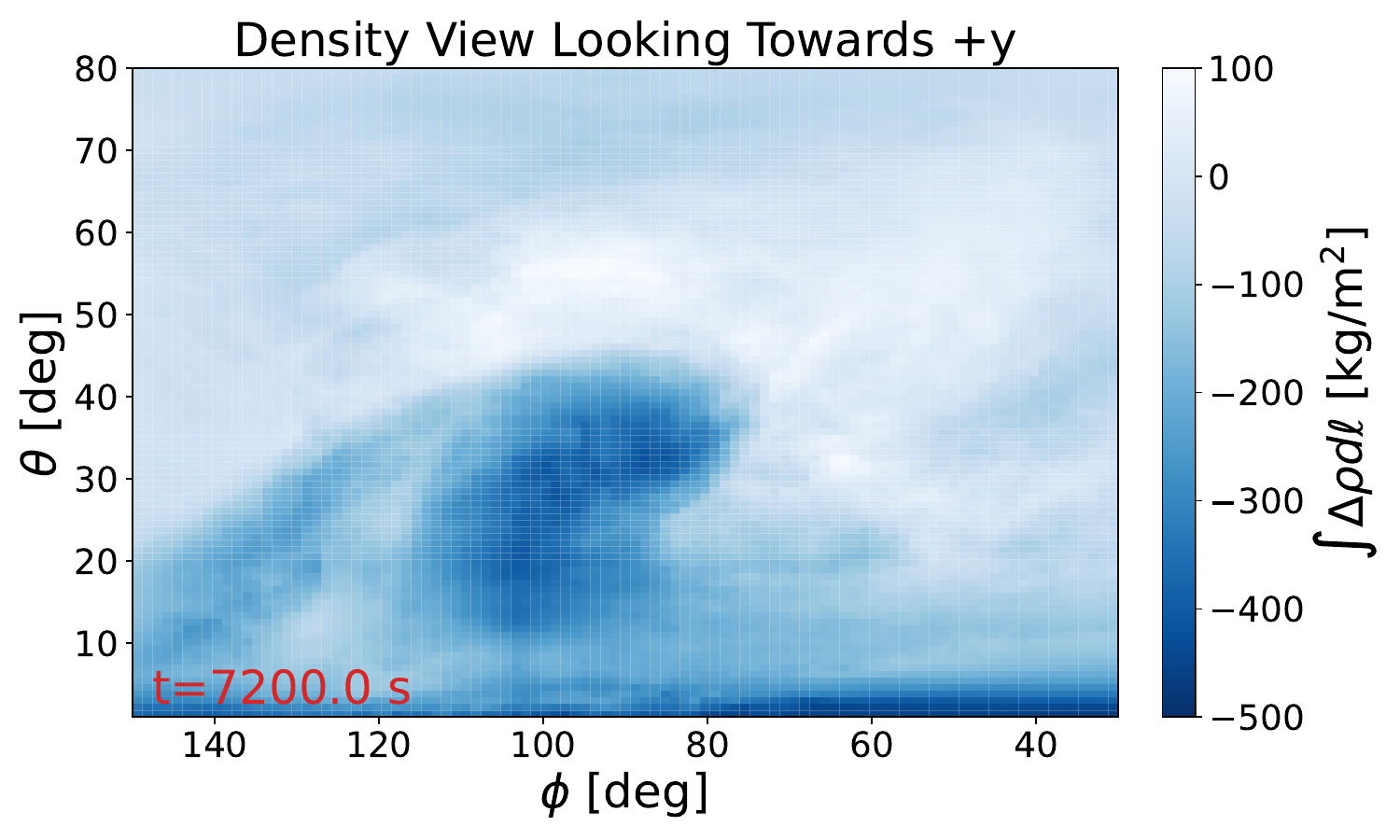}
    \includegraphics[width=0.4\textwidth]{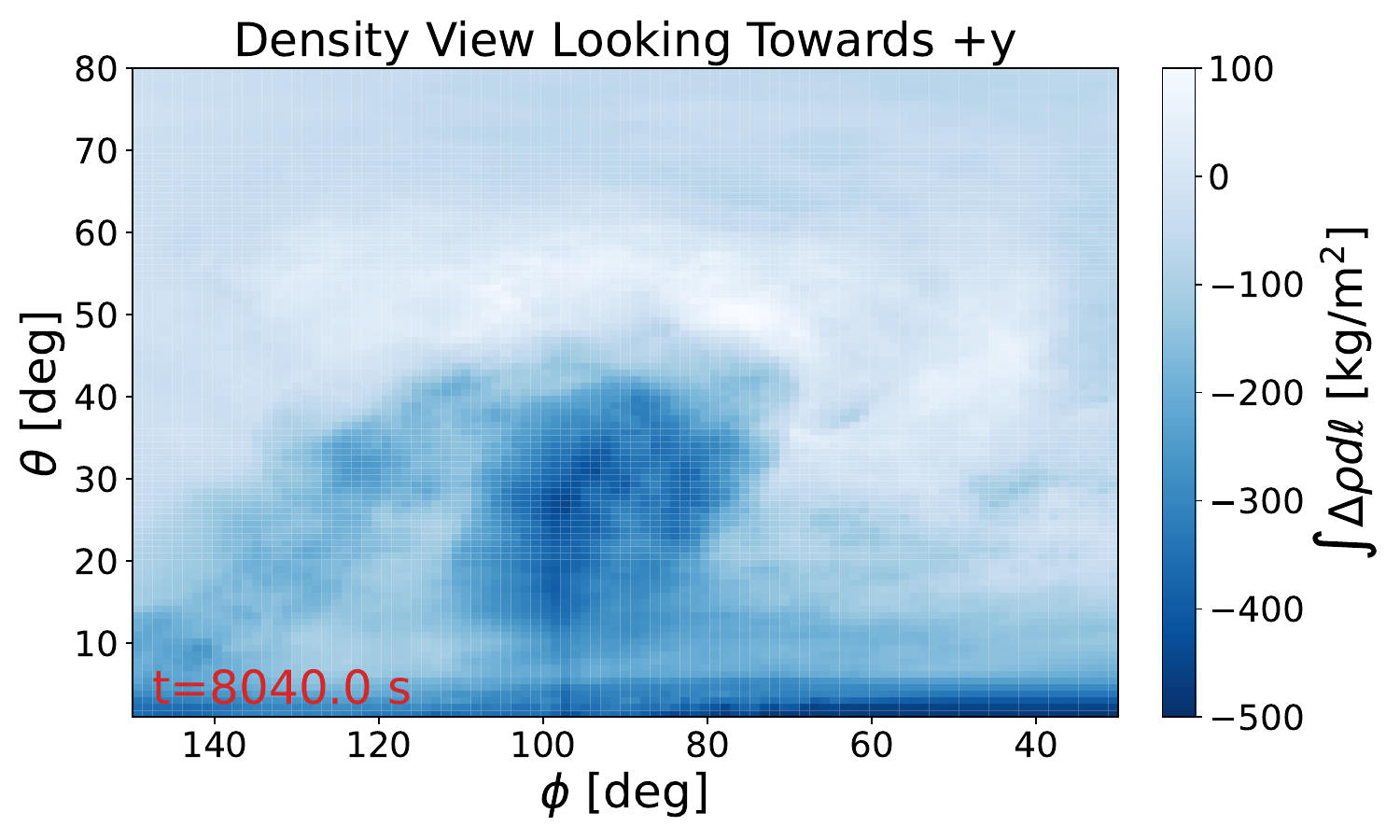}
    \includegraphics[width=0.4\textwidth]{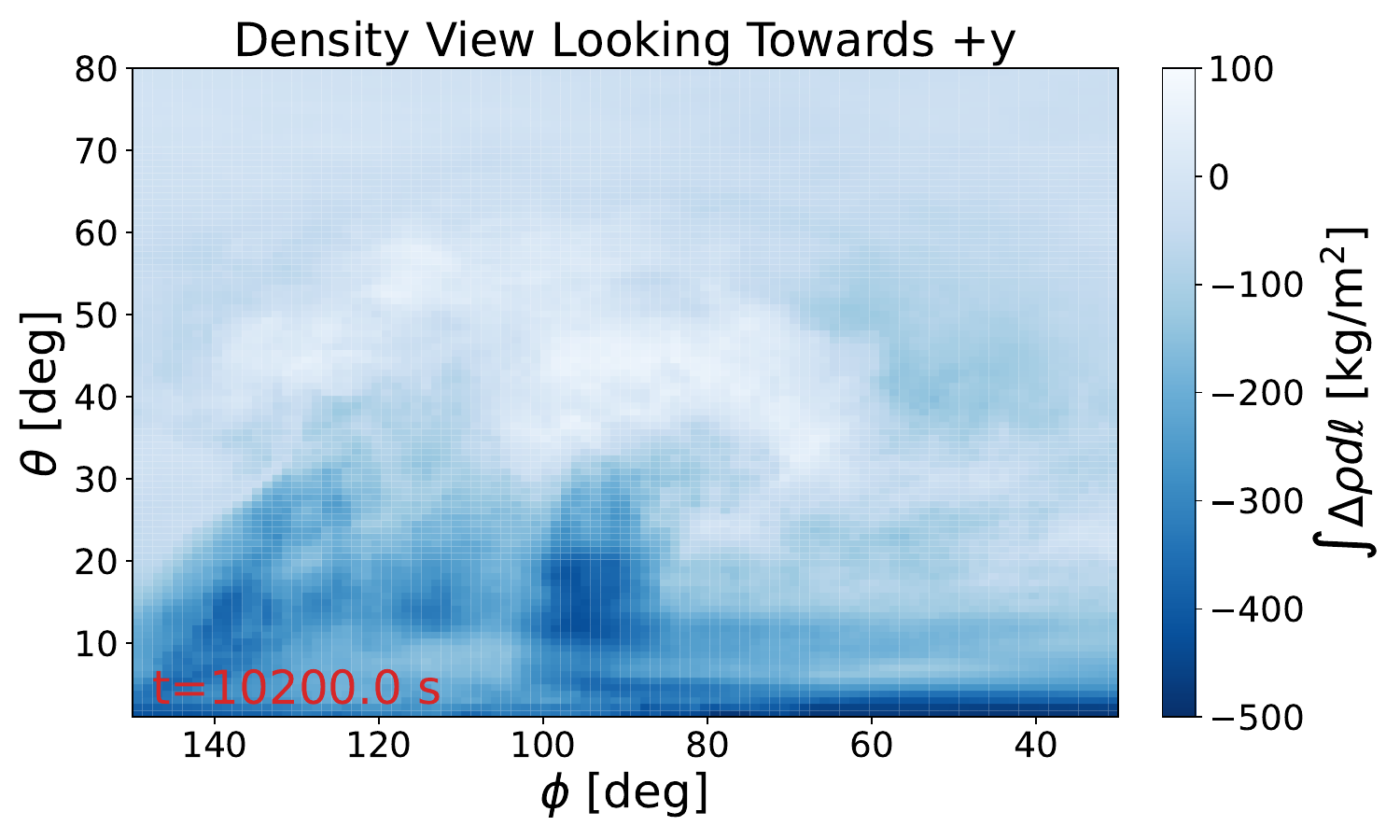}
\caption{\label{fig:integrateddensity}The density perturbation field, relative to the no-storm base state, integrated along line of sight leading to the muon detector (here, $d \ell$ is the line element along the line of sight), at various times during the simulated storm evolution. The tornado forms at $t=5580$ seconds, and is located within the mesocyclone producing the density perturbation near $\phi=100$ degrees.}
\end{figure*}

For comparison, plots showing the density perturbation field integrated along lines of sight leading to the muon detector are shown in figure \ref{fig:integrateddensity}. The large region of low density located near $\phi=100$ degrees corresponds to the storm's mesocyclone. While the mesocyclone is present during all stages of the storm's lifecycle, the intensification of this region between $t=5580$ and $t=10200$ seconds corresponds to the presence of a tornado. This intensification is reflected in the muon flux, as can be seen in plots of the idealized muon flux in figure \ref{fig:muonflux}. 

\begin{figure*}[t]
        \includegraphics[width=0.4\textwidth]{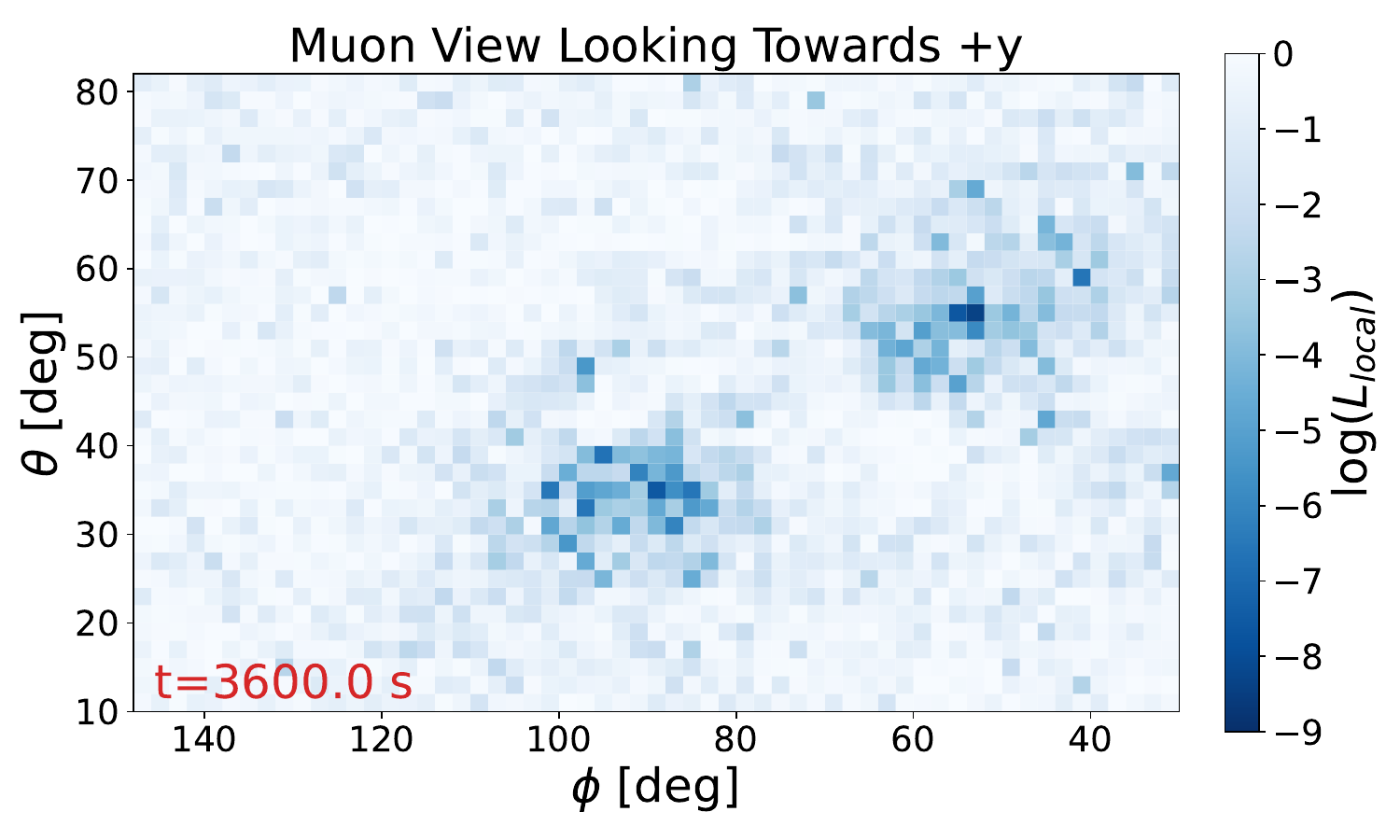}
        \includegraphics[width=0.4\textwidth]{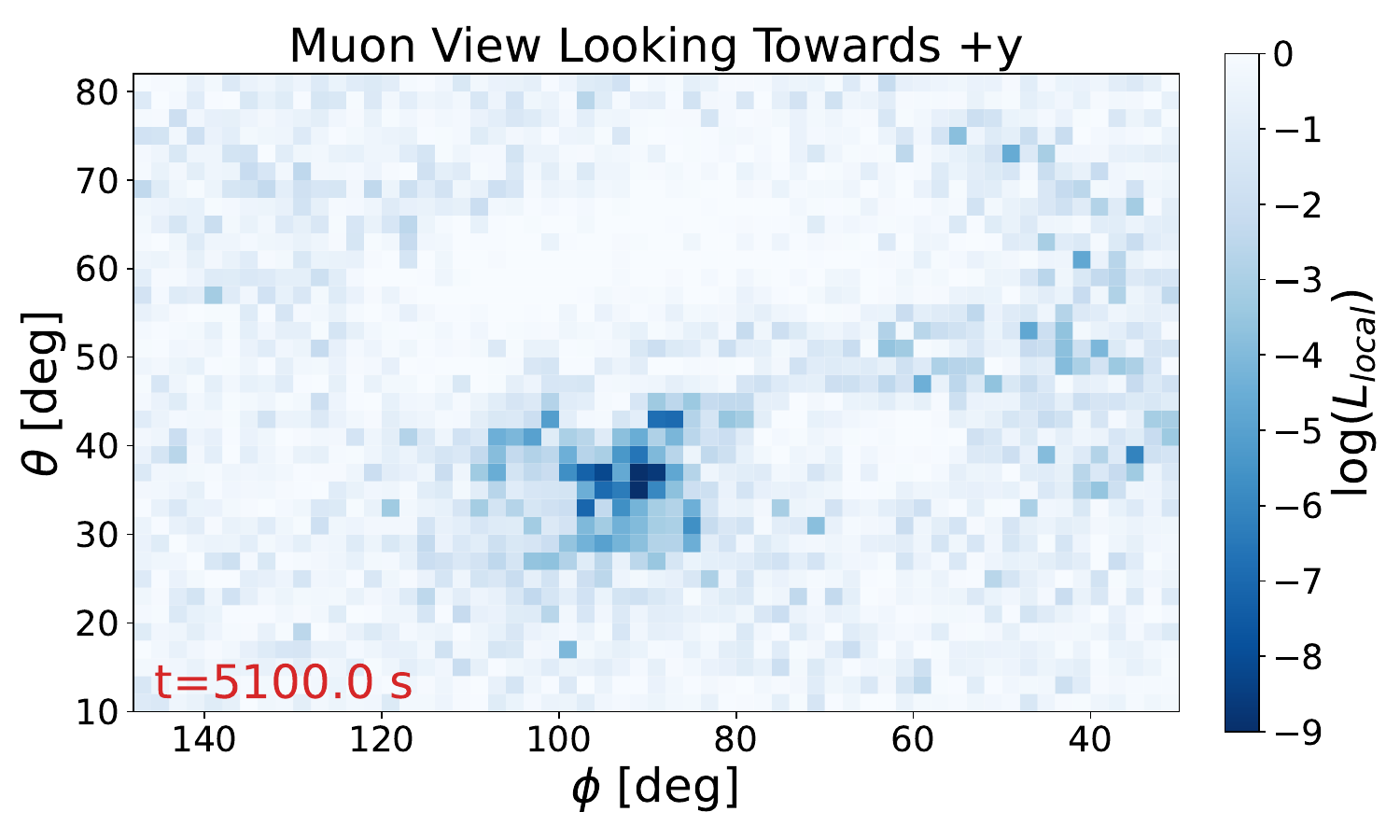}
        \includegraphics[width=0.4\textwidth]{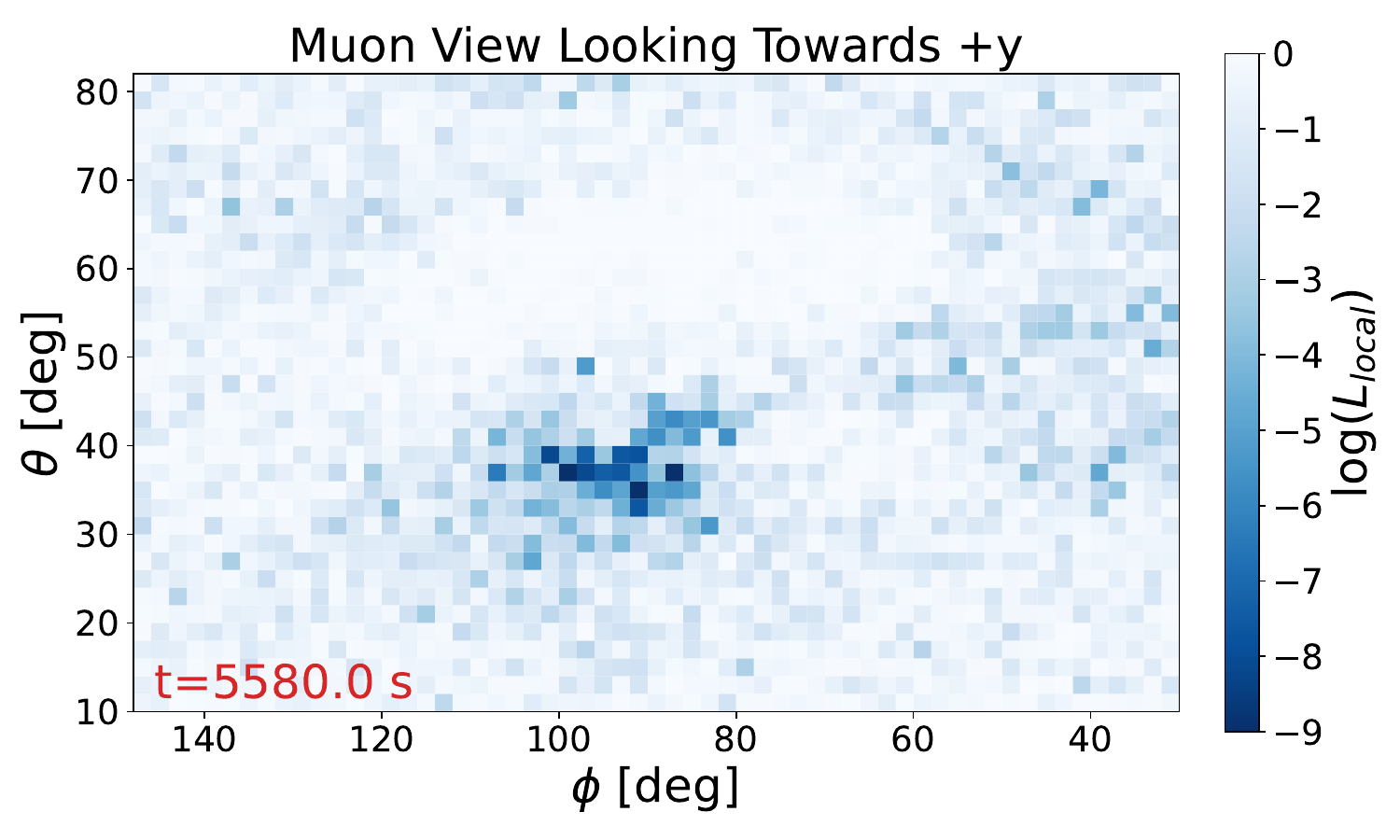}
        \includegraphics[width=0.4\textwidth]{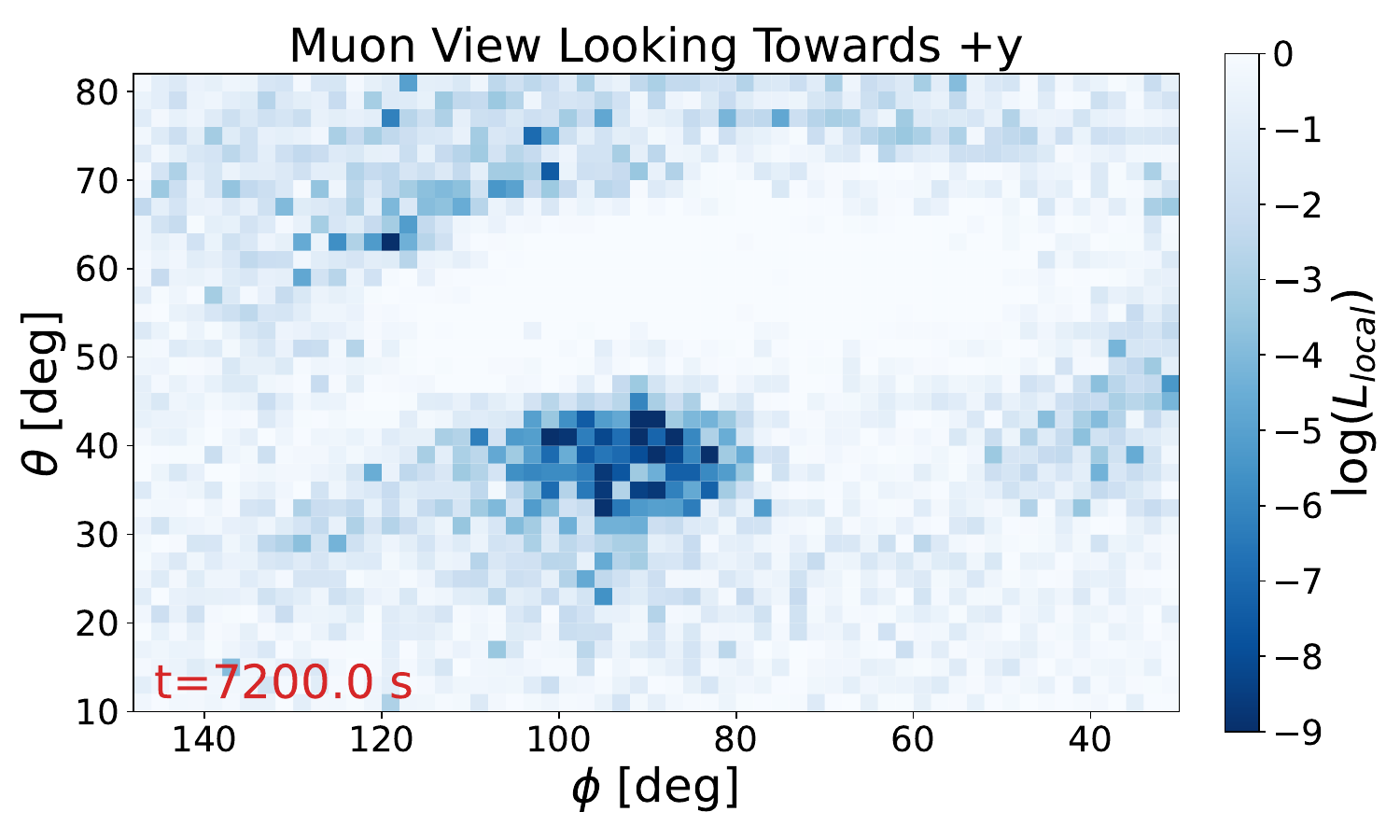}
        \includegraphics[width=0.4\textwidth]{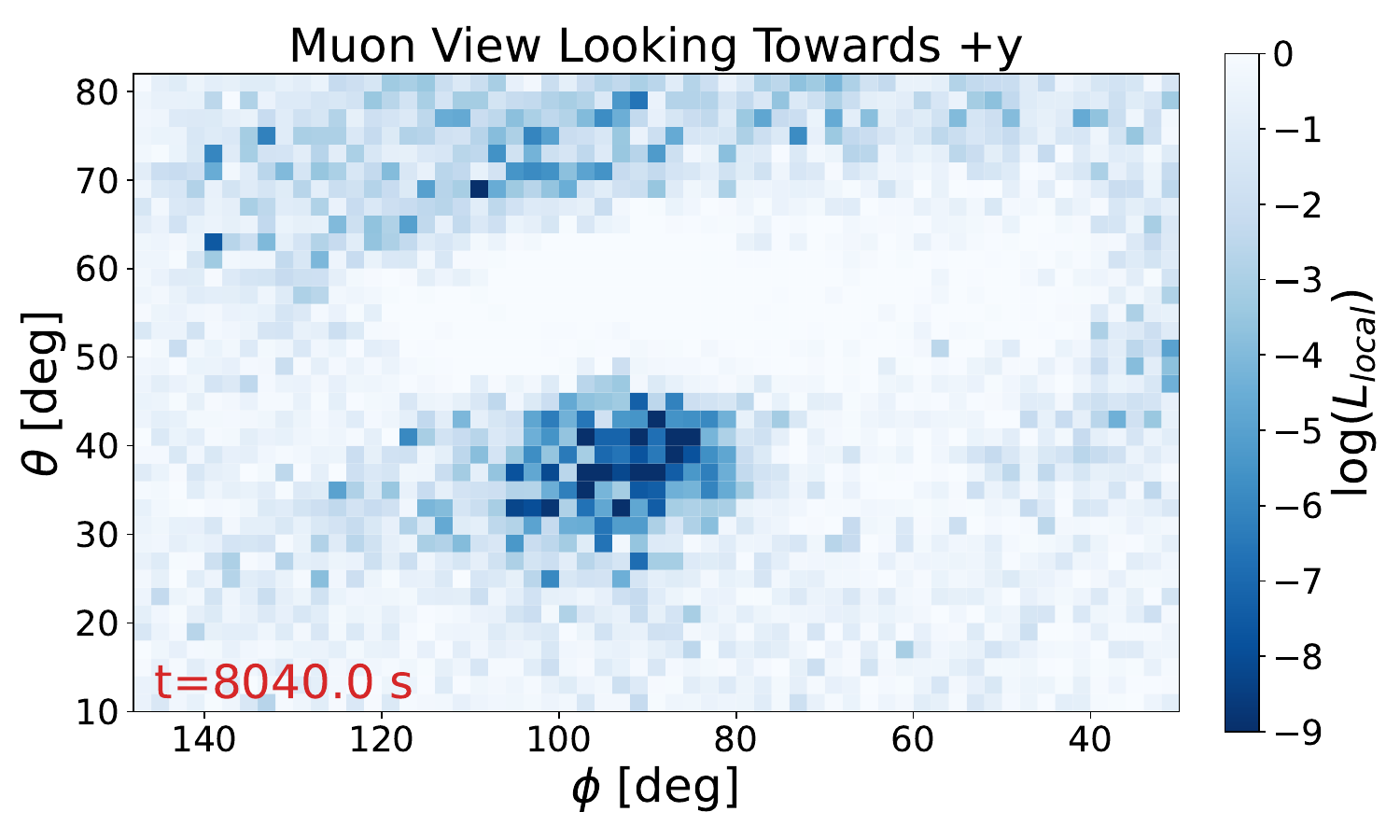}
        \includegraphics[width=0.4\textwidth]{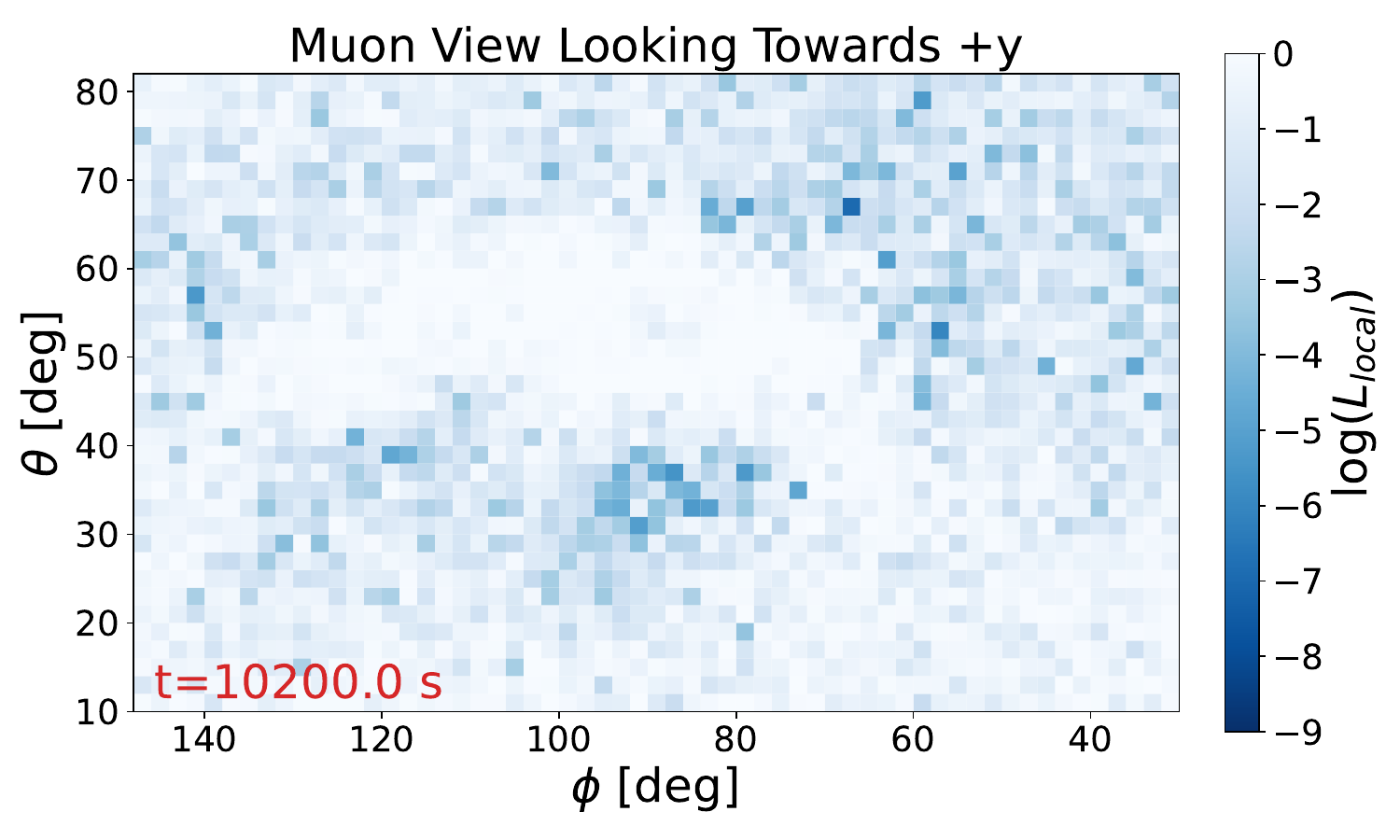}
    \caption{``Realistic" expectations of the effect of a supercell thunderstorm on the atmospheric muon flux, including statistical variations in the observed muon flux for a 1000 m$^2$ detector. Each image is obtained from integrating the muon flux over the preceding hour.}
    \label{fig:muonflux_poisson}
\end{figure*}

The plots show in figure \ref{fig:muonflux} show the idealized expectation of the effect of the simulated supercell thunderstorm on the atmospheric muon flux, neglecting statistical variations in the atmospheric muon flux. To improve upon this, Poisson variations are added to $N_{obs}$ and $L$ is recalculated using the new values. Examples of maps of $L$ containing these variations can be seen in figure \ref{fig:muonflux_poisson}. Near the horizon, the muon flux is lower and the data is dominated by Poisson variations. At high elevations ($\theta > 30$ degrees) the structure of the high and low density regions of the storm begin to become visible as clusters of pixels with lower/elevated $L$ values. The intensification of the low density core temporally coincident with tornado formation is still visible, even when accounting for statistical variations in the observed muon flux over a 1 hour time window. 

\begin{figure}[h]
    \centering
    \includegraphics[width=0.4\textwidth]{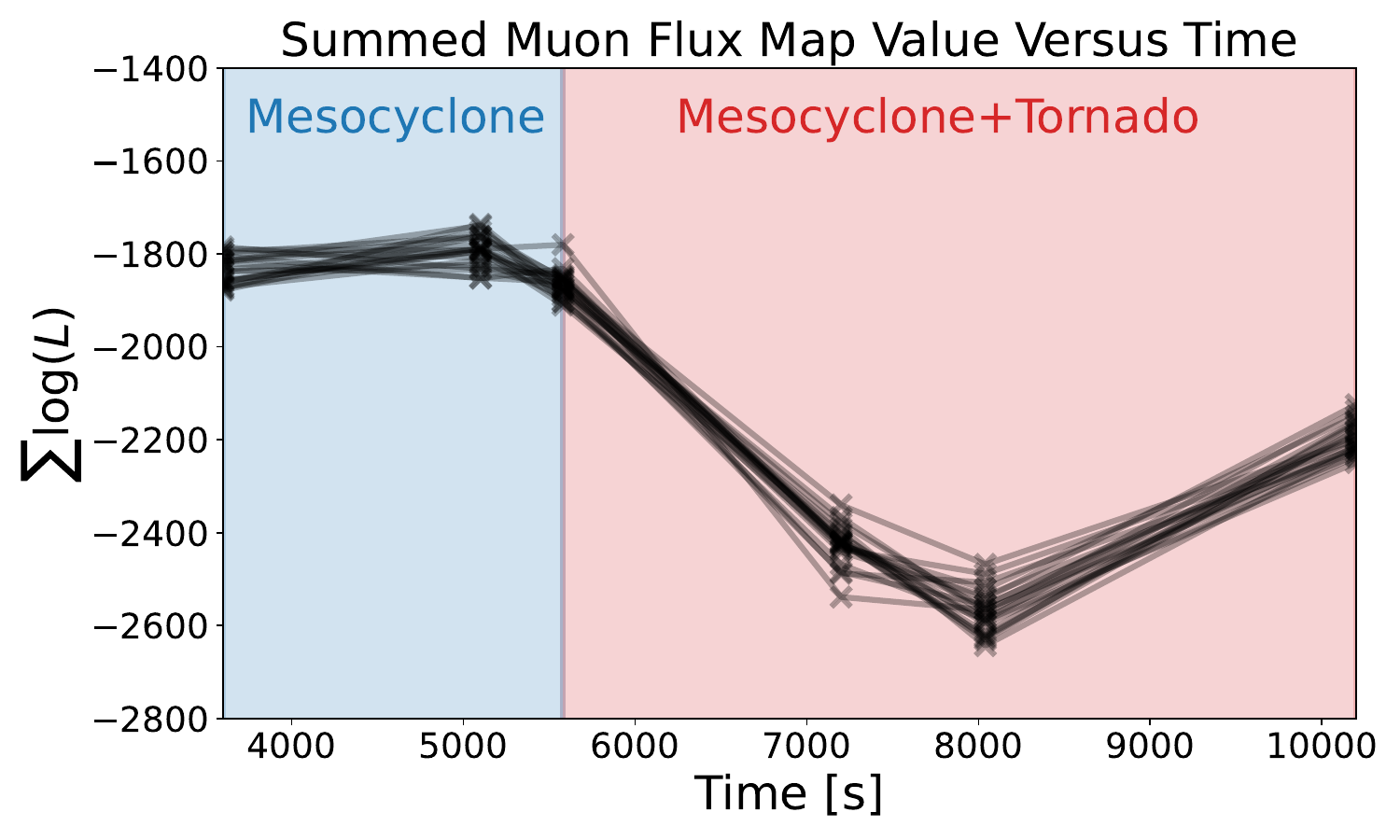}
    \caption{The sum of the pixel values of each of the plots in figure \ref{fig:muonflux_poisson}, plotted for various points in the storm's life cycle. 20 realizations of muon data associated with this storm are shown to give an estimate of the statistical spread that could be expected from a muon measurement. The effect of the tornado can be seen as a decrease in the $\sum \log L$ value following tornado formation at $t=5580$ seconds. Note that this decrease in in addition to the existing deficit due to the presence of the larger mesocyclone.}
    \label{fig:muflux_v_t}
\end{figure}

The effect of the tornado can also be seen in the total summed value of the maps shown in figure \ref{fig:muonflux_poisson}. Low density perturbations will introduce an excess of muon events seen by the detector, resulting in lower per-pixel $\log(L)$ values on average. Figure \ref{fig:muflux_v_t} shows the value of the summed muon images as a function of time during the supercell thunderstorm's life span. Tornado formation introduces a drop in the spatially integrated muon flux beyond the effect seen from the mesocyclone alone. 

\section{Discussion}
\subsection{Muon Detector Size}
The previous studies presented in this paper use a fairly large 1000 m$^2$ muon detector as a baseline. There are several existing cosmic ray detectors of this scale or larger that are sensitive to the atmospheric muon flux~\cite{Auger}\cite{Teshima:1997sc}, however these detectors are not currently capable of reconstructing the incoming directions of incoming single muons. Additionally, there have (un)fortunately been no recorded tornadoes in the vicinity of either of these detectors, as they are not located in particularly tornado-active regions~\cite{TornadoTracks}. Given that tornado detection often relies on eyewitness observation, and that these detectors are in remote locations, it is possible that tornadoes occurred near these detectors but were not officially recorded. If this is the case, an investigation into historical muon data from these detectors could reveal a signature corresponding to a previously unidentified tornado, if the tornado were intense enough to cause a significant decrease in the total atmospheric muon flux. 

\begin{figure*}
    \includegraphics[width=0.4\textwidth]{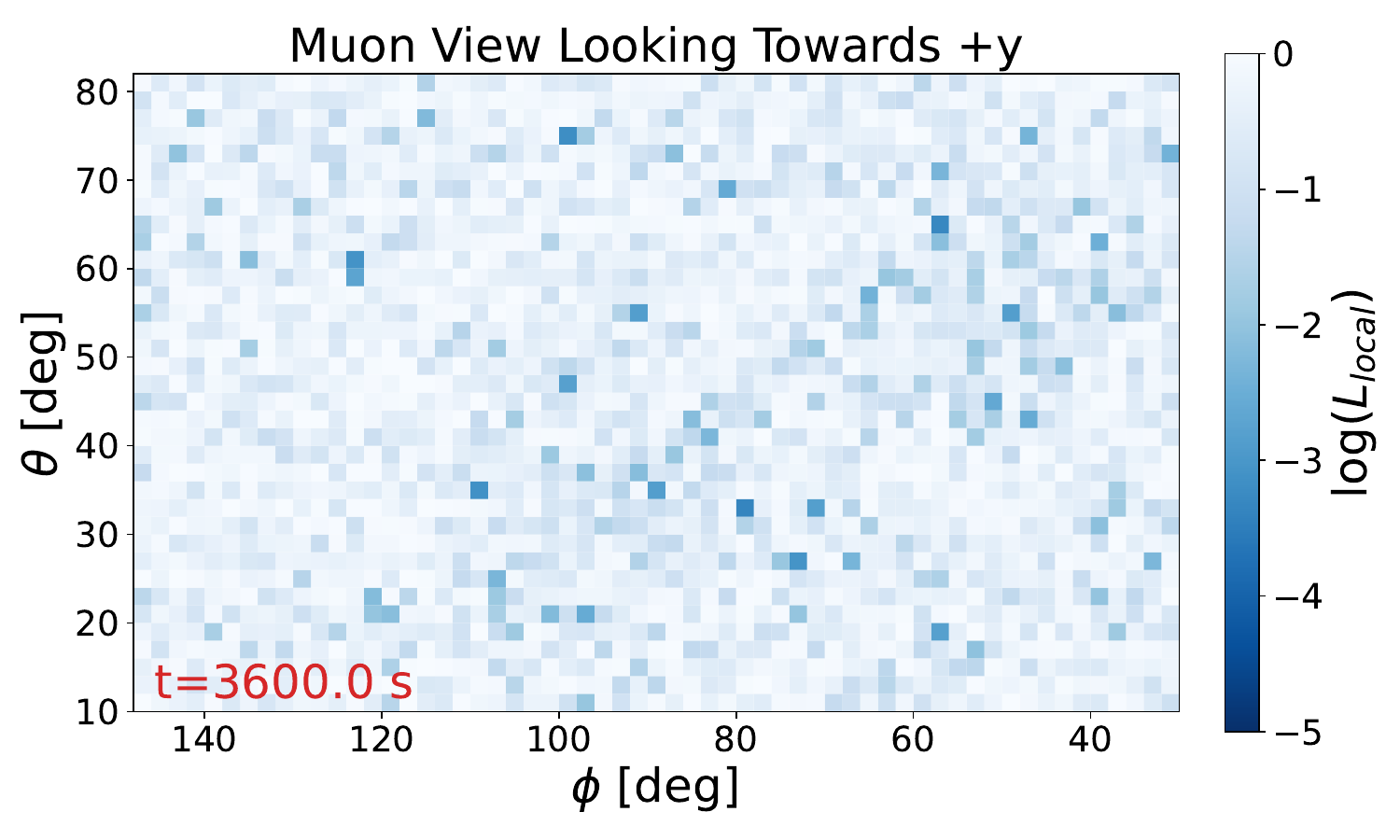}
    \includegraphics[width=0.4\textwidth]{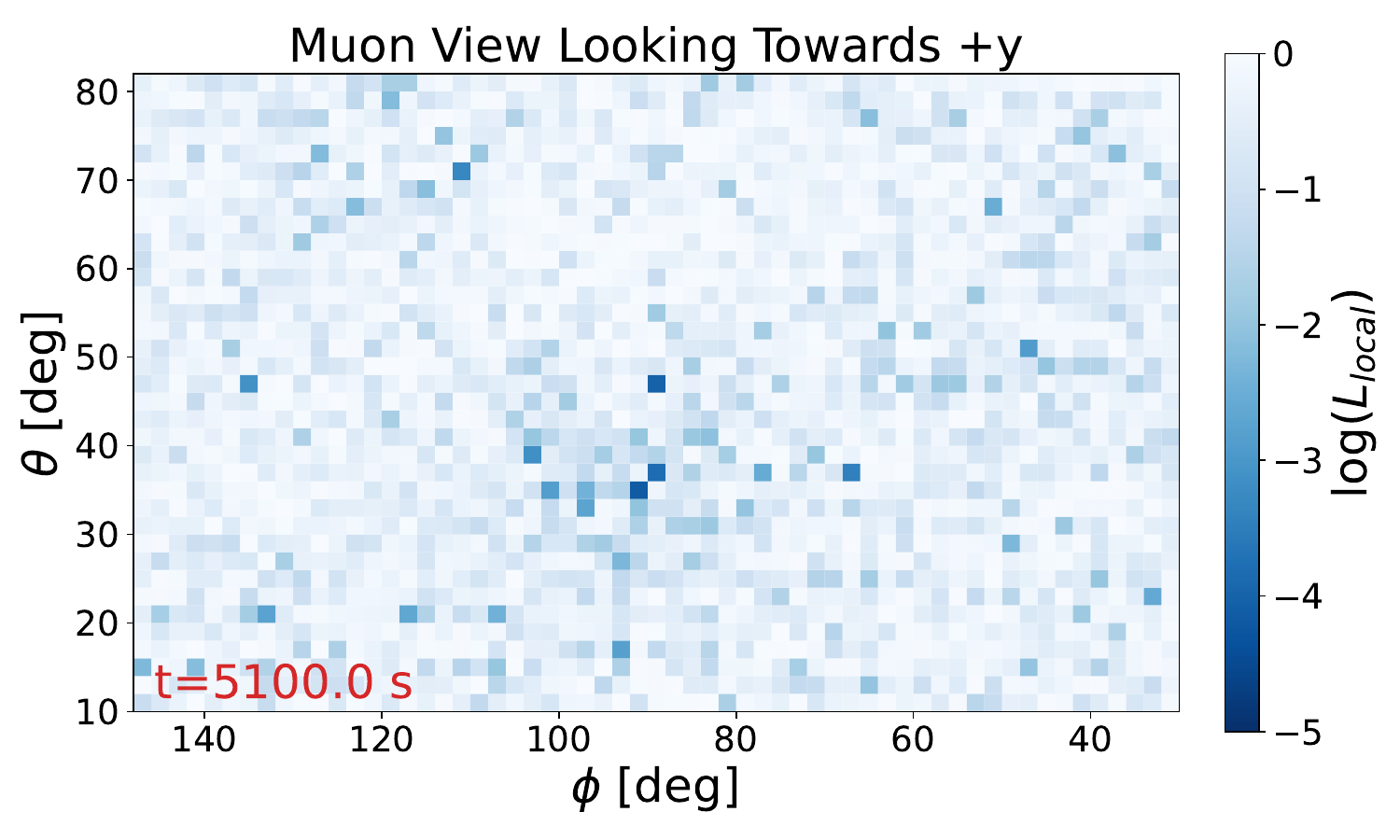}
    \includegraphics[width=0.4\textwidth]{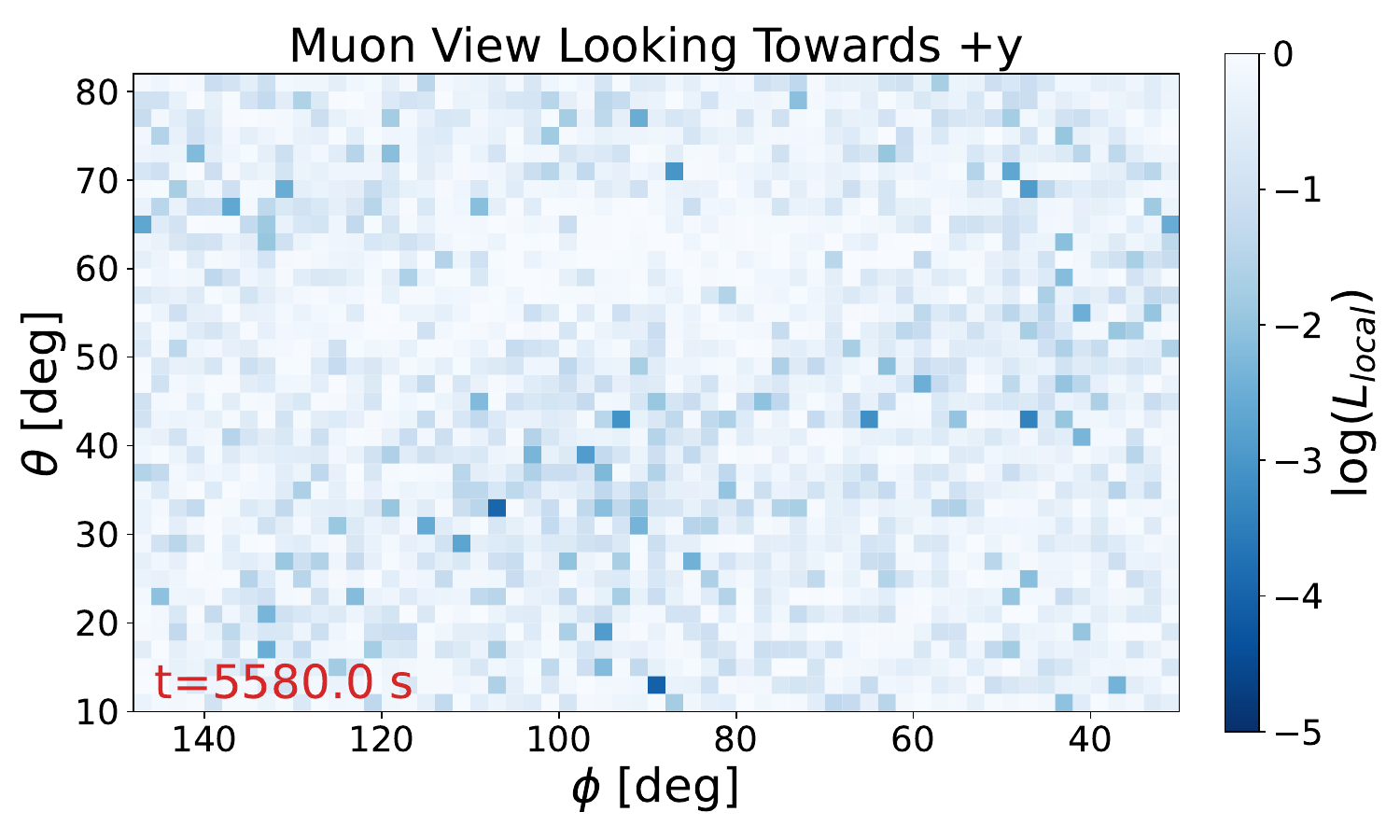}
    \includegraphics[width=0.4\textwidth]{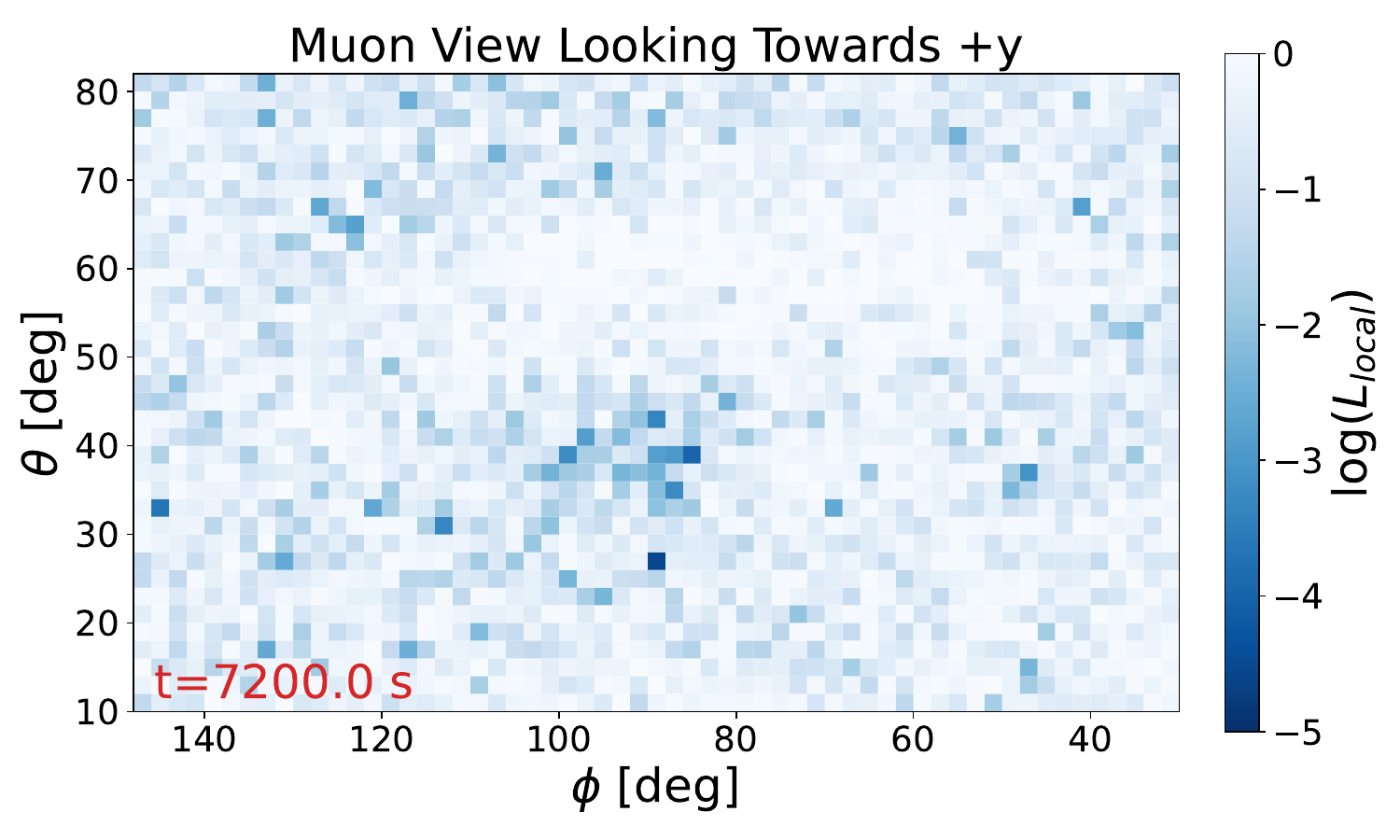}
    \includegraphics[width=0.4\textwidth]{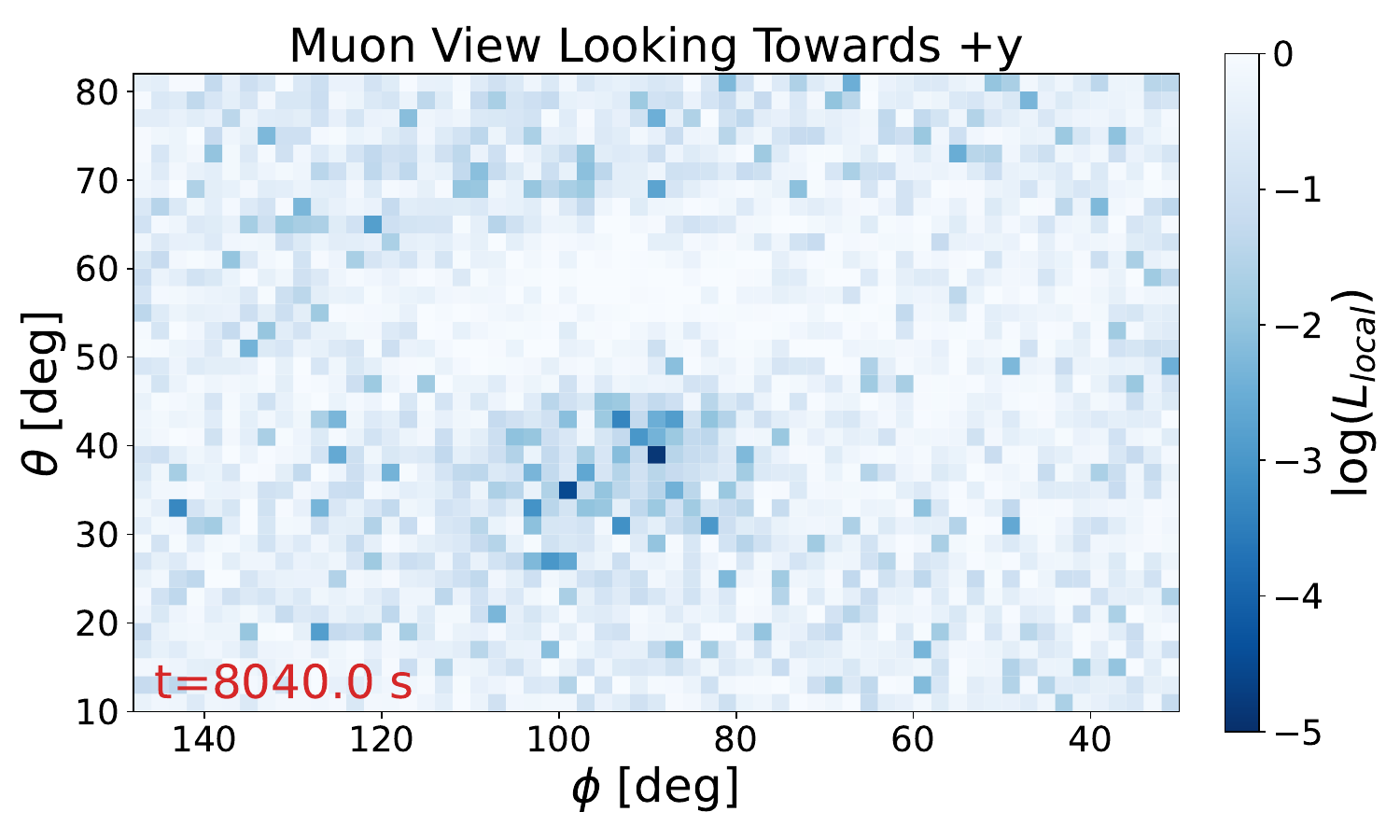}
    \includegraphics[width=0.4\textwidth]{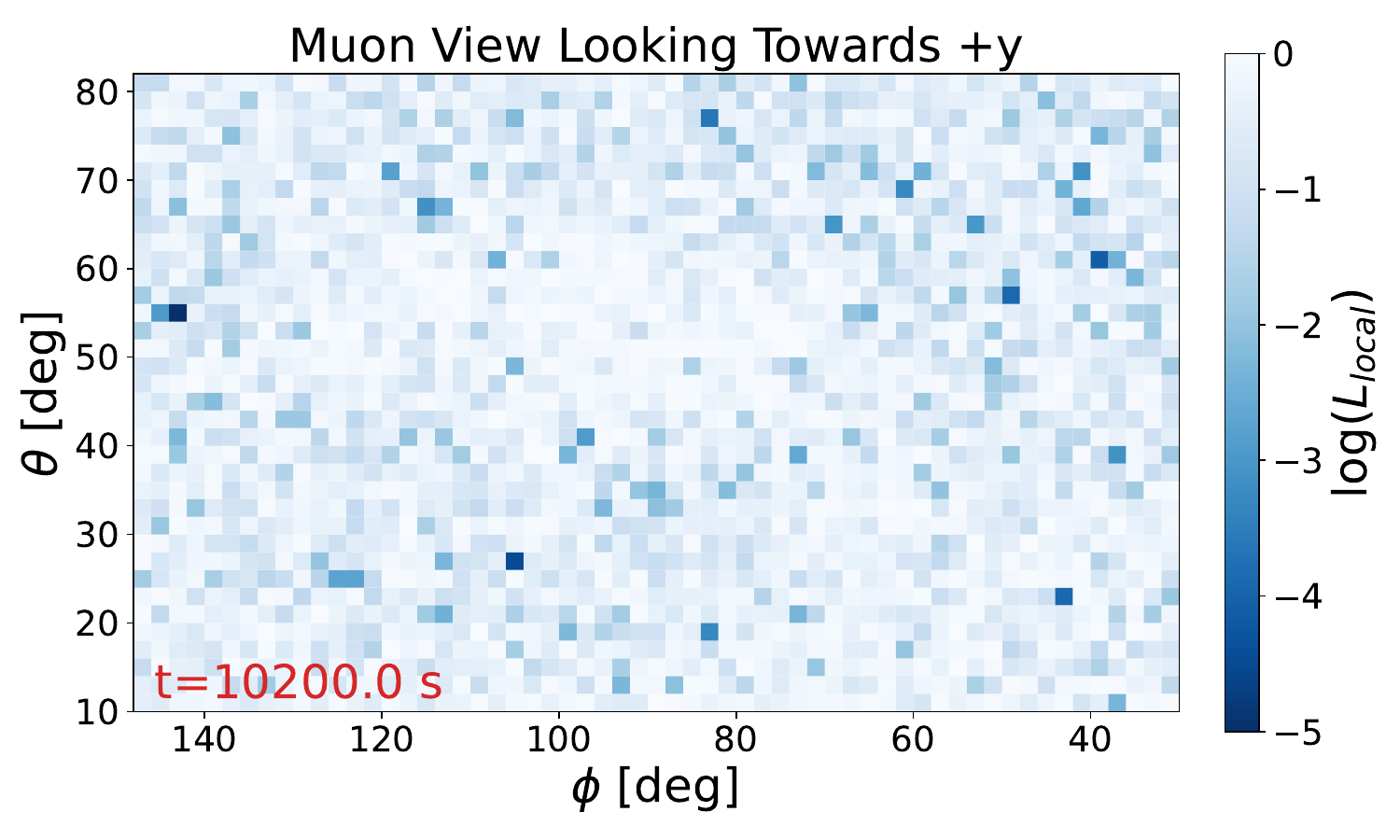}
    \caption{Maps of the muon flux, similar to those shown in figure \ref{fig:muonflux_poisson}, but as seen by a 100 m$^2$ muon detector. The effect of the low-density core of the tornado is significantly smaller relative to statistical fluctuations in comparison to the 1000 m$^2$ case, though is still faintly visible.}
    \label{fig:smalldetmaps}
\end{figure*}

\begin{figure}
    \centering
    \includegraphics[width=0.4\textwidth]{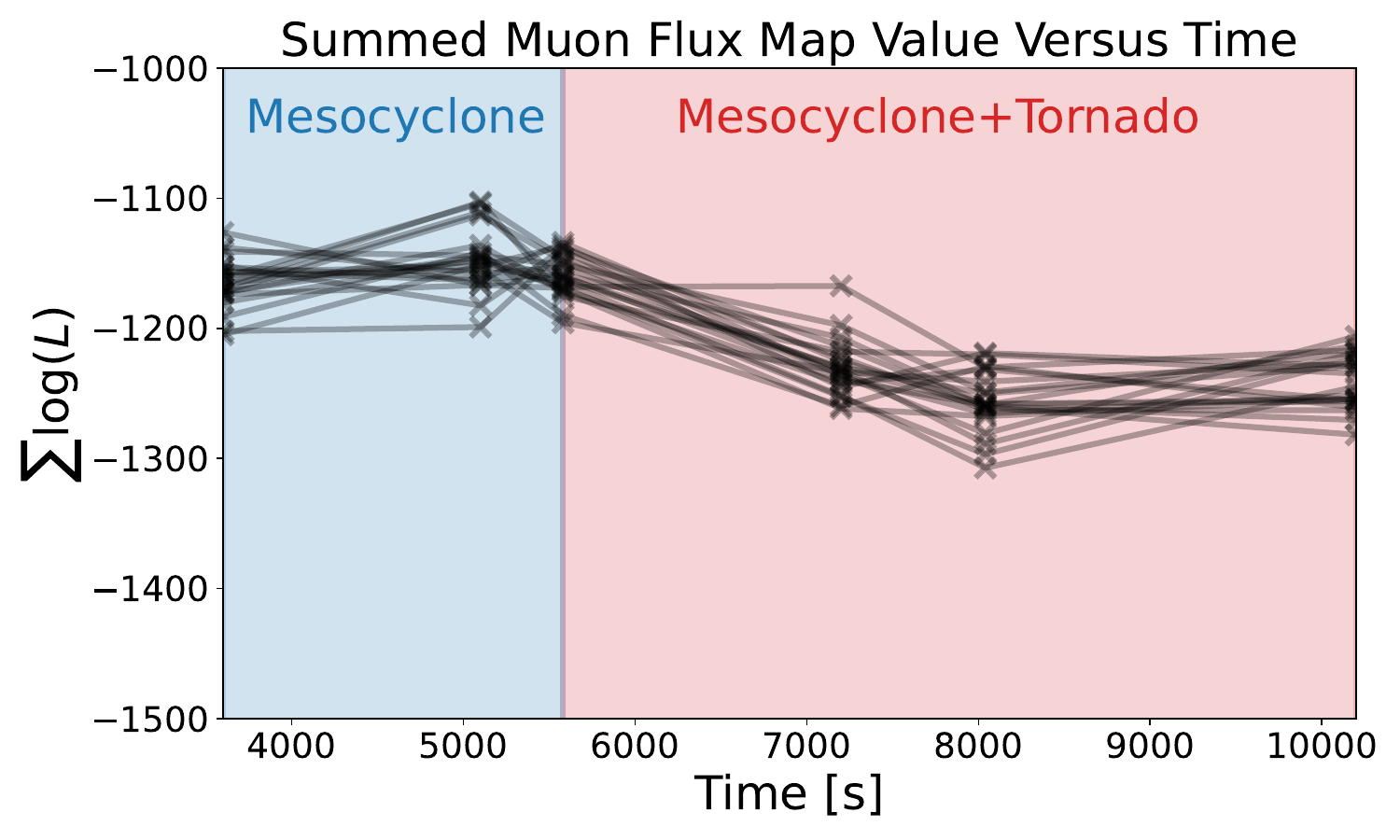}
    \caption{The sum of the pixel values of each of the plots in figure \ref{fig:smalldetmaps}, plotted for various points in the storm's life cycle, as seen by a 100 m$^2$ muon detector. 20 realizations of muon data associated with this storm are shown to give an estimate of the statistical spread that could be expected from a muon measurement. While the effect of tornado formation is smaller than that seen in figure \ref{fig:muflux_v_t}, a decrease in $\sum \log L$ values is still clearly visible following tornado formation.}
    \label{fig:smalldetsum}
\end{figure}

Smaller detectors may still be able to measure atmospheric density effects on the muon flux. Since variations in the atmospheric muon flux due to supercell thunderstorms are on the order of 1\%, muon detectors seeking to measure this effect need to be large enough to collect O($10^6$) atmospheric muon counts over the angular area of interest during the lifetime of the storm (typically 10s to 100s of minutes). Measurements taken with smaller detectors will have larger relative statistical fluctuations, making it more difficult to identify the underlying structure of the density field. On the other hand, smaller detectors also have the potential to be significantly more portable. While a 1000 m$^2$ would likely need to be a permanent, stationary installation, a 100 m$^2$ detector could conceivably be moved to a location with projected severe weather several days beforehand, and a 50 m$^2$ detector is small enough to fit on roughly 5 trucks, allowing for the detector to be moved to the location of a severe weather event as it is developing. Examples of muon maps produced by a 100 m$^2$ detector can be seen in figure \ref{fig:smalldetmaps}. Though statistical variations conceal much of the structure seen in figure \ref{fig:muonflux_poisson}, the aggregate effect of the tornado on the underlying density field can still be observed in the sum of the individual pixel values, albeit to a much smaller degree. This is shown in figure \ref{fig:smalldetsum}. 

\subsection{Range and Resolution}
Though individual density measurements using this technique may not be as precise in measuring the density at a particular location when compared to more traditional in-situ methods, there is a distinct advantage in being able to remotely measure the density field over large distance scales. Measurements of the atmospheric muon flux can reveal unique information of the variation of the density field as a function of height that is inaccessible to ground-based sensors. Additionally, the lack of a need for in-situ instrumentation makes this measurement more logistically straightforward than current methods, as it does not require detectors to be within the direct path of an active tornado. 

The range of this technique depends on the desired spatial resolution of the density perturbation field. The atmospheric muon flux naturally varies with elevation angle, with a higher proportion of muon events incident at large elevation angles~\cite{GaisserCR}.  For a fixed detector size, a more detailed scan of the density field requires a larger number of muon counts, which can be obtained by using the muon flux a higher (more vertical) elevation angles. However, using data at higher elevation angles reduces the maximum possible horizontal distance between the detector and the storm. 

\begin{figure}
    \centering
    \includegraphics[width=0.4\textwidth]{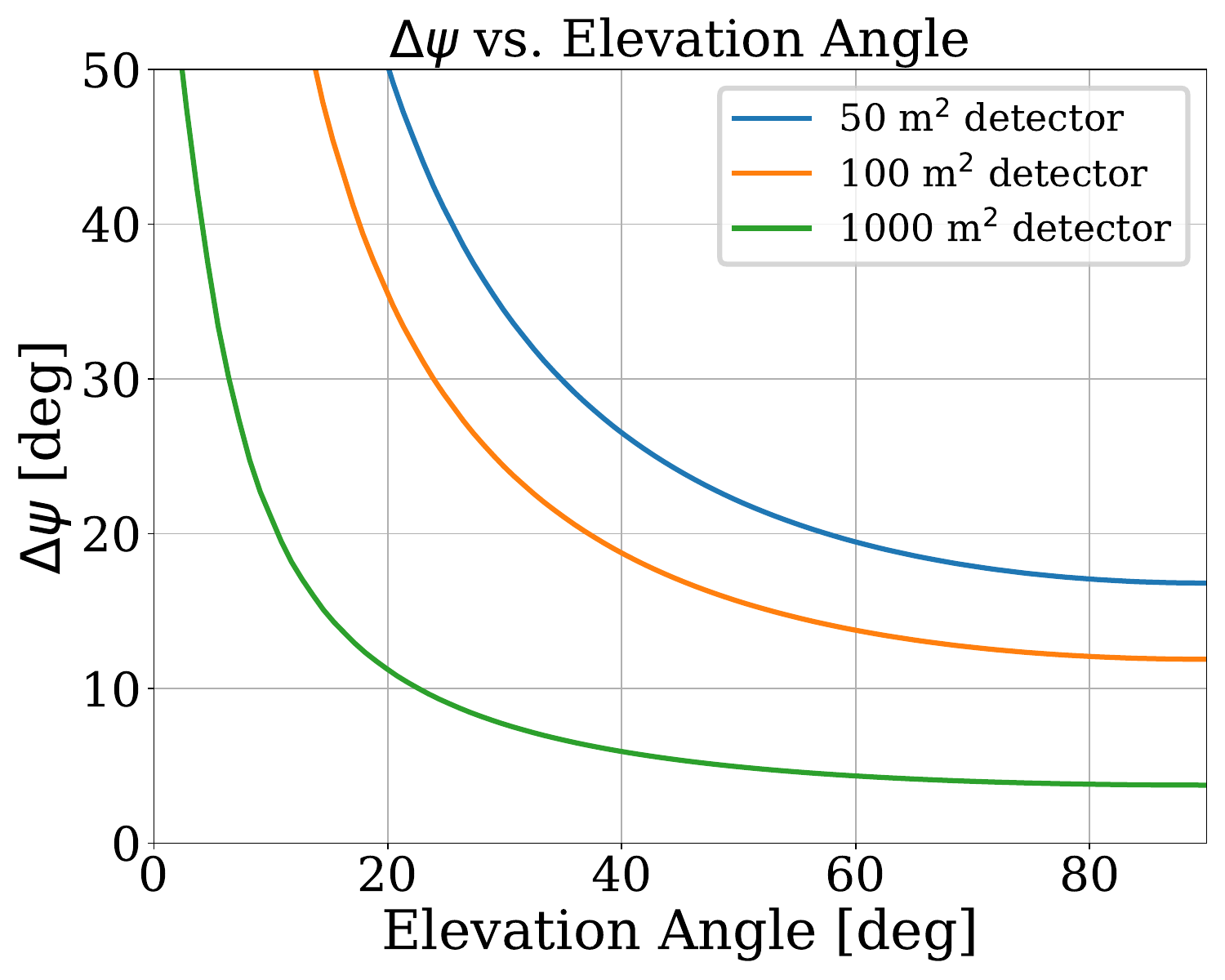}
    \caption{The angular bin size required to capture $10^6$ muons at various elevations angles and various detector sizes. Since the muon flux decreases for small elevation angles, larger angular bins are needed to capture a fixed number of muons. This translates to concealing variations in the muon flux on small length scales in these regions, as there is an insufficient density of muons to measure these effects.}
    \label{fig:resoplot}
\end{figure}

\begin{figure}
    \includegraphics[width=0.4\textwidth]{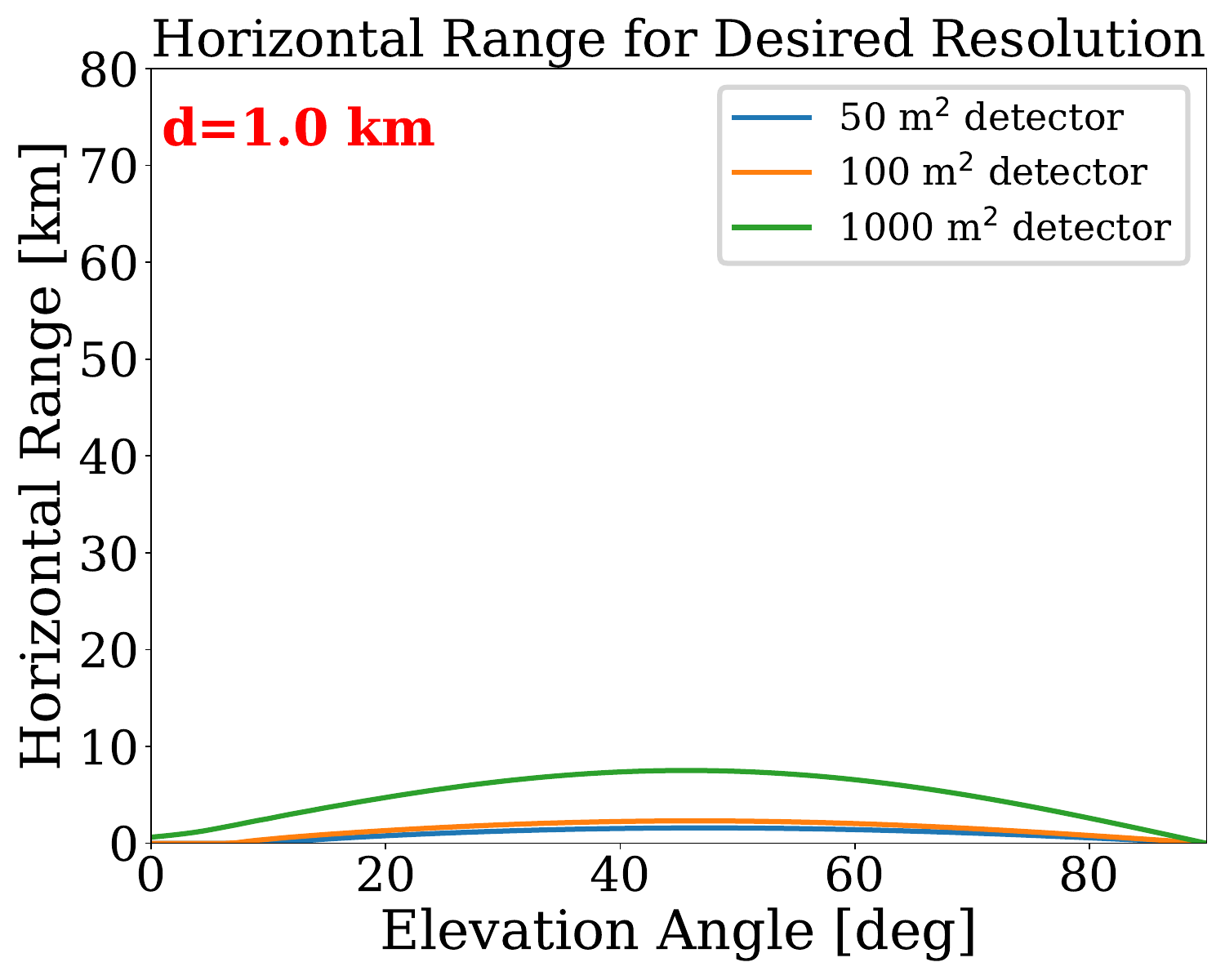}
    \includegraphics[width=0.4\textwidth]{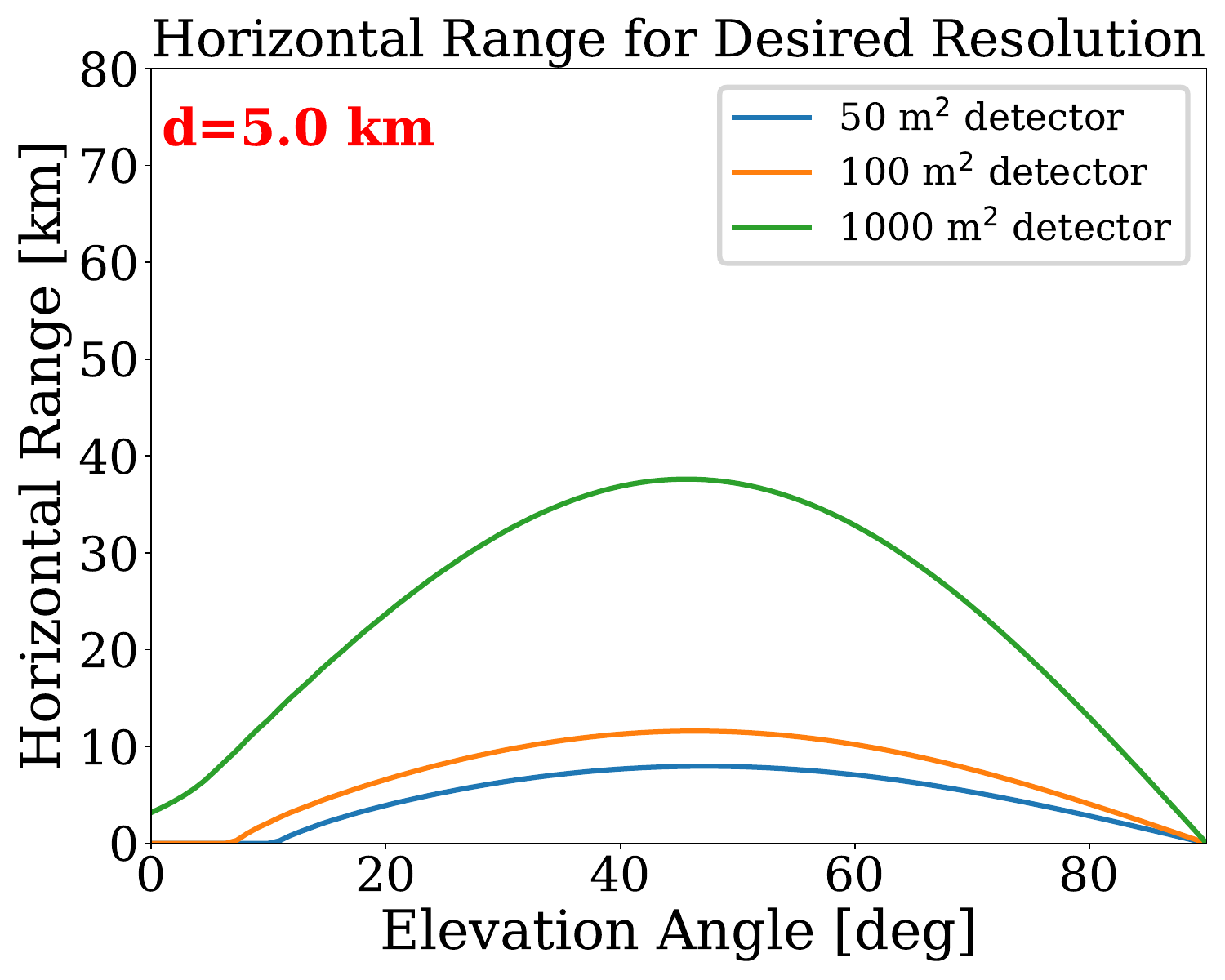}
    \includegraphics[width=0.4\textwidth]{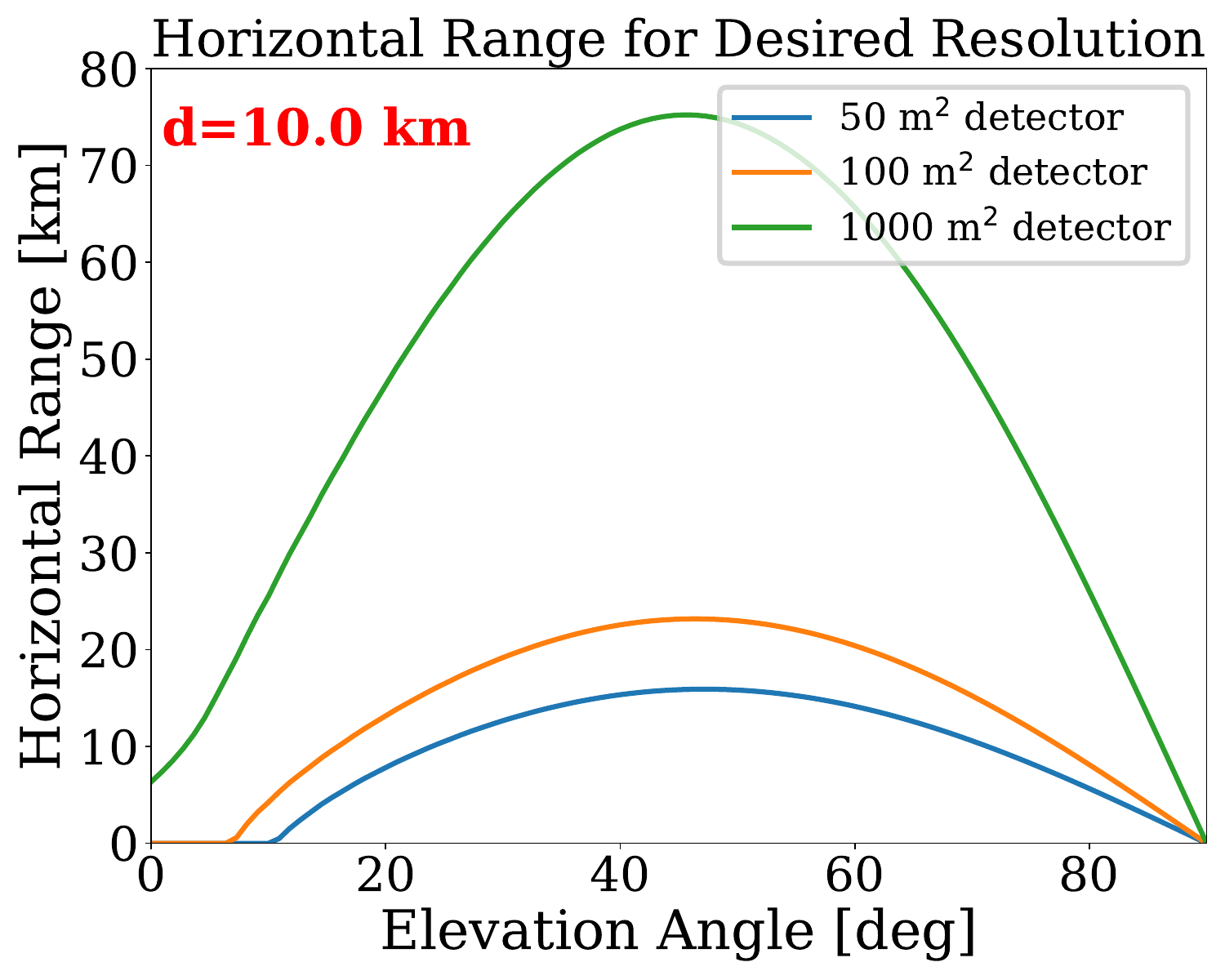}
    \caption{Projected horizontal ranges of using this technique to measure density field variations along the listed length scales ($d$). Large elevation angles correspond to a higher muon rate allowing for more precise measurements, but also geometrically correspond to a smaller horizontal distance to a particular altitude layer. These two competing effects are roughly balanced near an elevation angle of 45 degrees, where the horizontal range of this technique is maximized.}
    \label{fig:rangeplot}
\end{figure}

To calculate the resolution of this technique, we require an observation of $10^6$ events within an angular radius of $\Delta\psi$ degrees (the ``resolution" of this technique). We assume that the error on reconstructing the directions of individual muon events is small relative to $\Delta\psi$, though in practice this will depend on the specifics of the detector. We can compute the range of this technique ($R$) for a particular desired distance scale and elevation angle with the following geometric relation:

\begin{equation}
    R = R_{Earth} \times \text{arctan} \left[ \frac{d}{\text{tan}(\Delta\psi)} \frac{\text{cos}(\theta)}{R_{Earth}} \right]
\end{equation}

where $d$ is the distance scale over which density variations are measured, $\theta$ is the elevation angle of the target region, and $\Delta\psi$ is obtained for a particular elevation angle from the curves shown in figure \ref{fig:resoplot}. Due to the competing effects of the increased resolution at low elevation angles, and the decreased horizontal distance when examining high elevation angles, the horizontal range of this technique peaks near an elevation angle of 45 degrees. This can be seen for various detector sizes and desired distance resolutions in figure \ref{fig:rangeplot}. Note that here ``horizontal range" is the distance along the surface of the earth, accounting for the earth's curvature. 

With a 1000 m$^2$ muon detector, a 5 kilometer resolution measurement of the density field could be obtained at a distance of up to 35 kilometers, though this is restricted to along the $\theta=40^{\circ}$ line of sight. If a larger, permanent muon detector were placed in a tornado-prone location, this would translate to being able to measure the density field of $O(1)$ tornadoes every year~\cite{TornadoTracks}. Smaller detectors would have a reduced, but still significant range, capable of measuring density variations over 5 kilometer length scales at a range of 10 kilometers. Such detectors would have the important advantage of portability, with the potential to be deployed or moved near regions expected to experience severe weather in the near future. 

\subsection{Hydrometeors}
\begin{figure*}
        \includegraphics[width=0.4\textwidth]{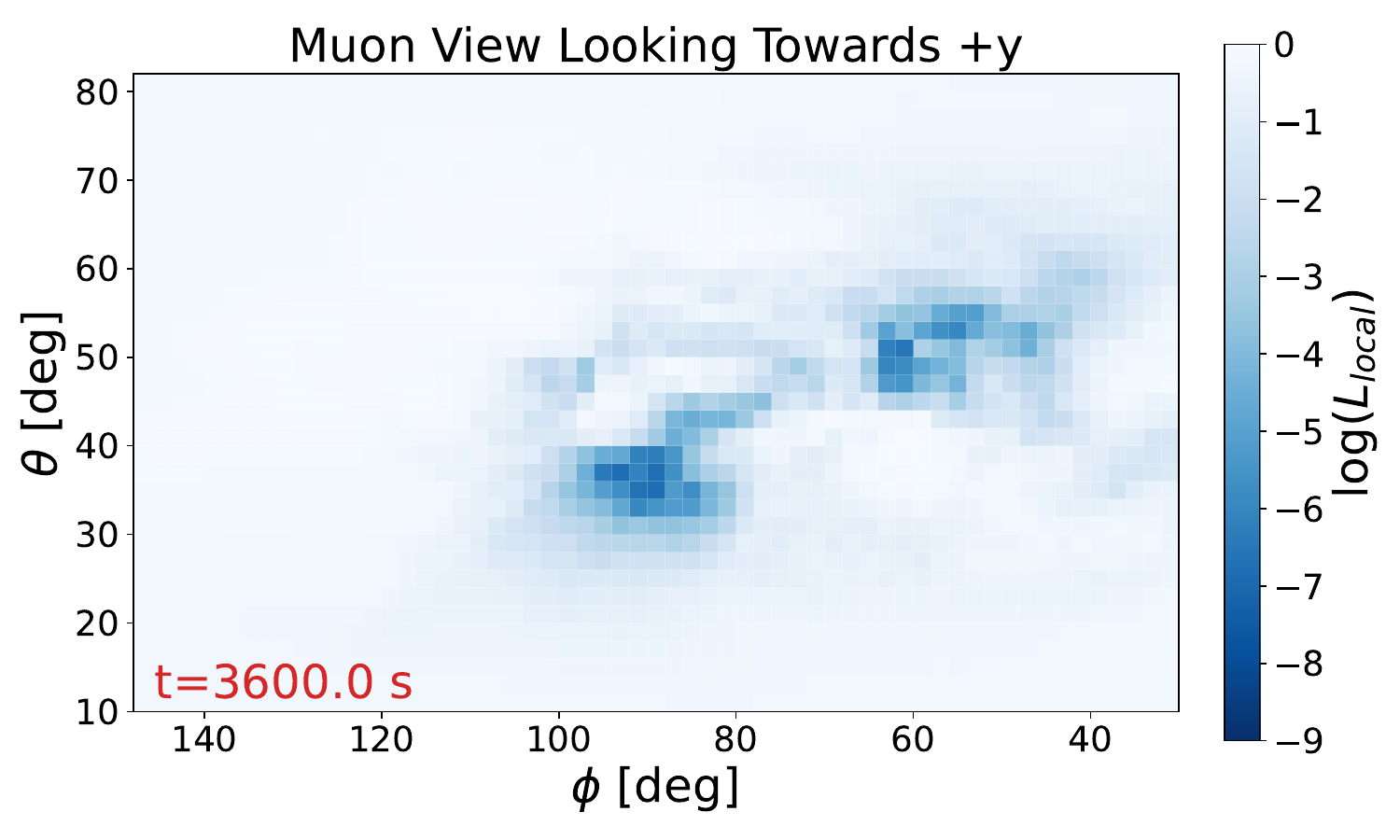}
        \includegraphics[width=0.4\textwidth]{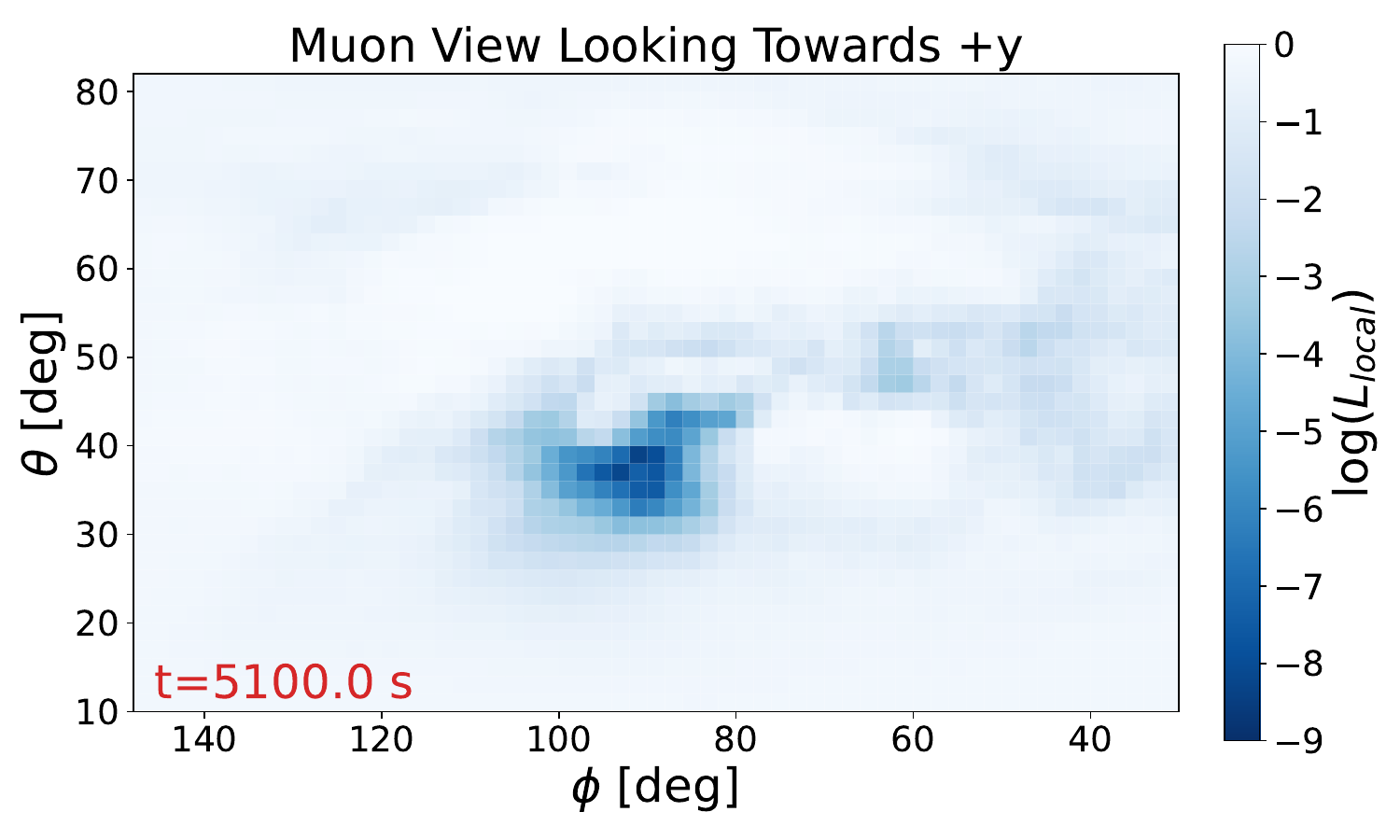}
        \includegraphics[width=0.4\textwidth]{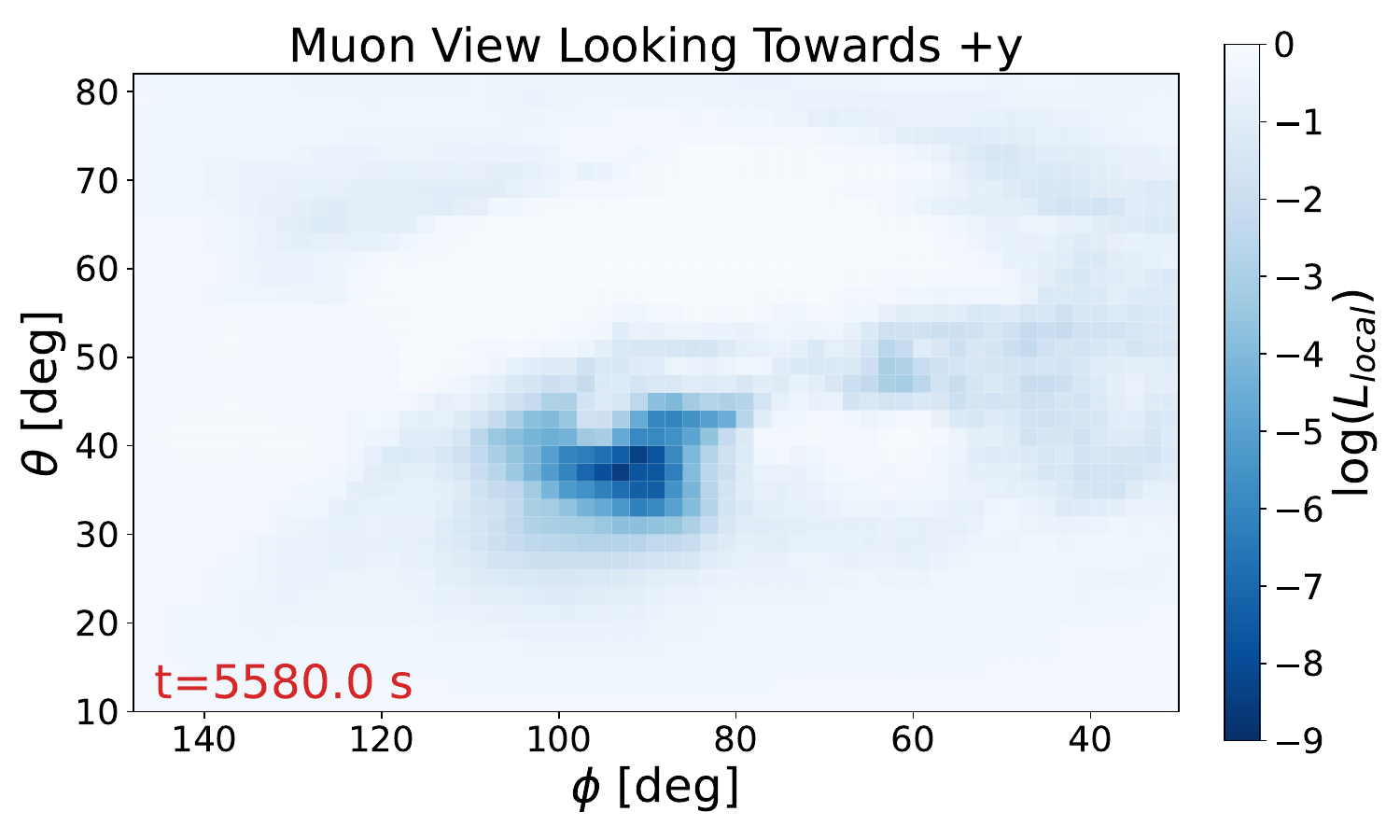}
        \includegraphics[width=0.4\textwidth]{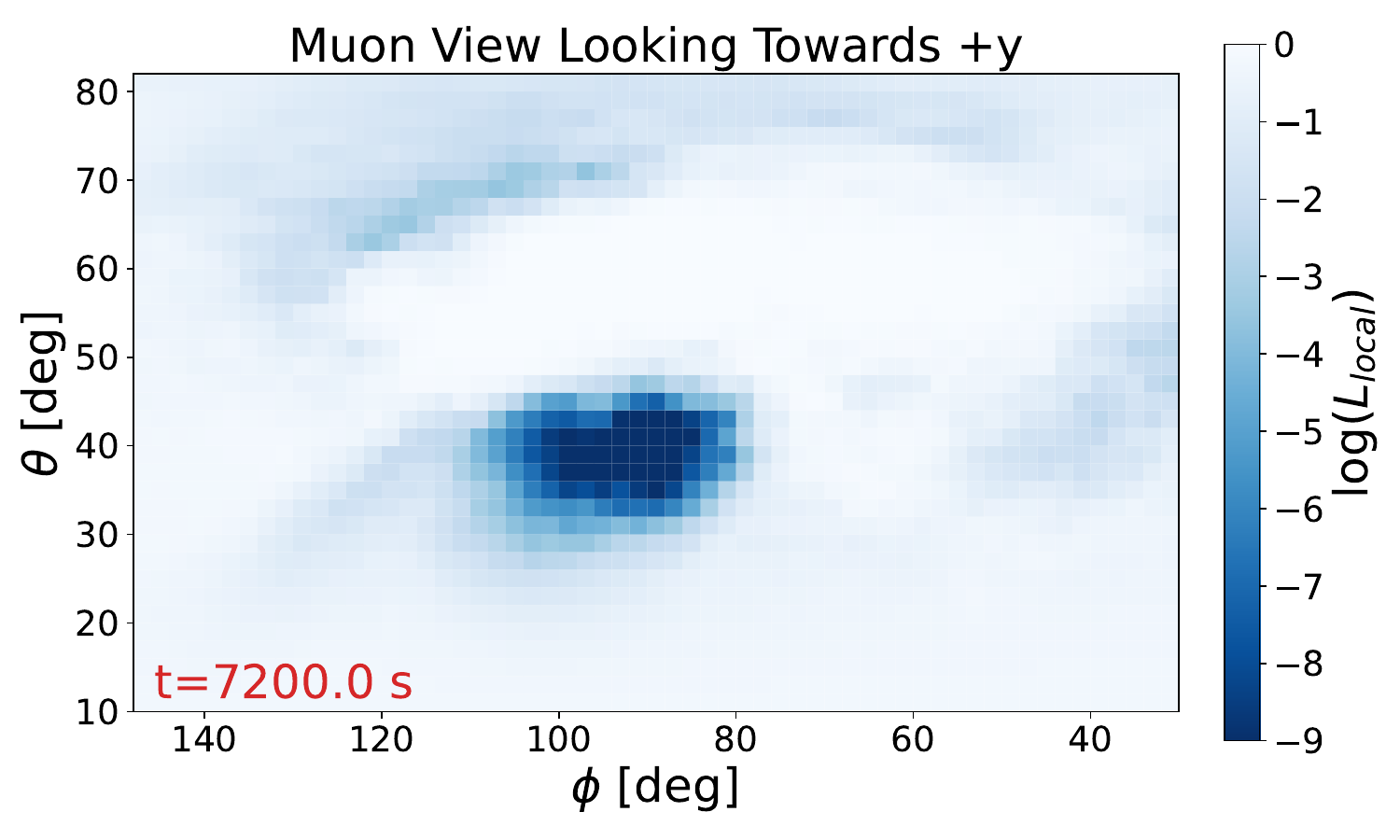}
        \includegraphics[width=0.4\textwidth]{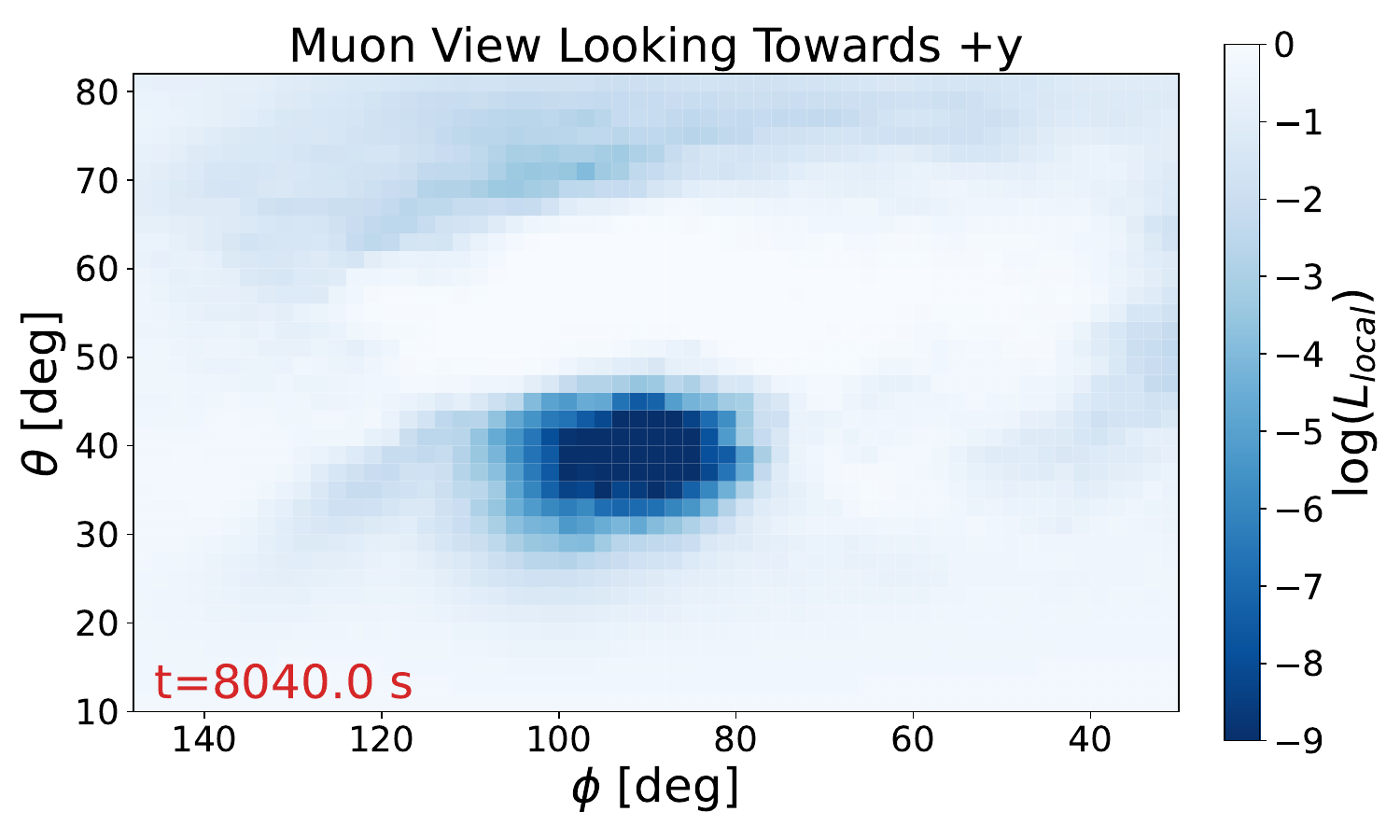}
        \includegraphics[width=0.4\textwidth]{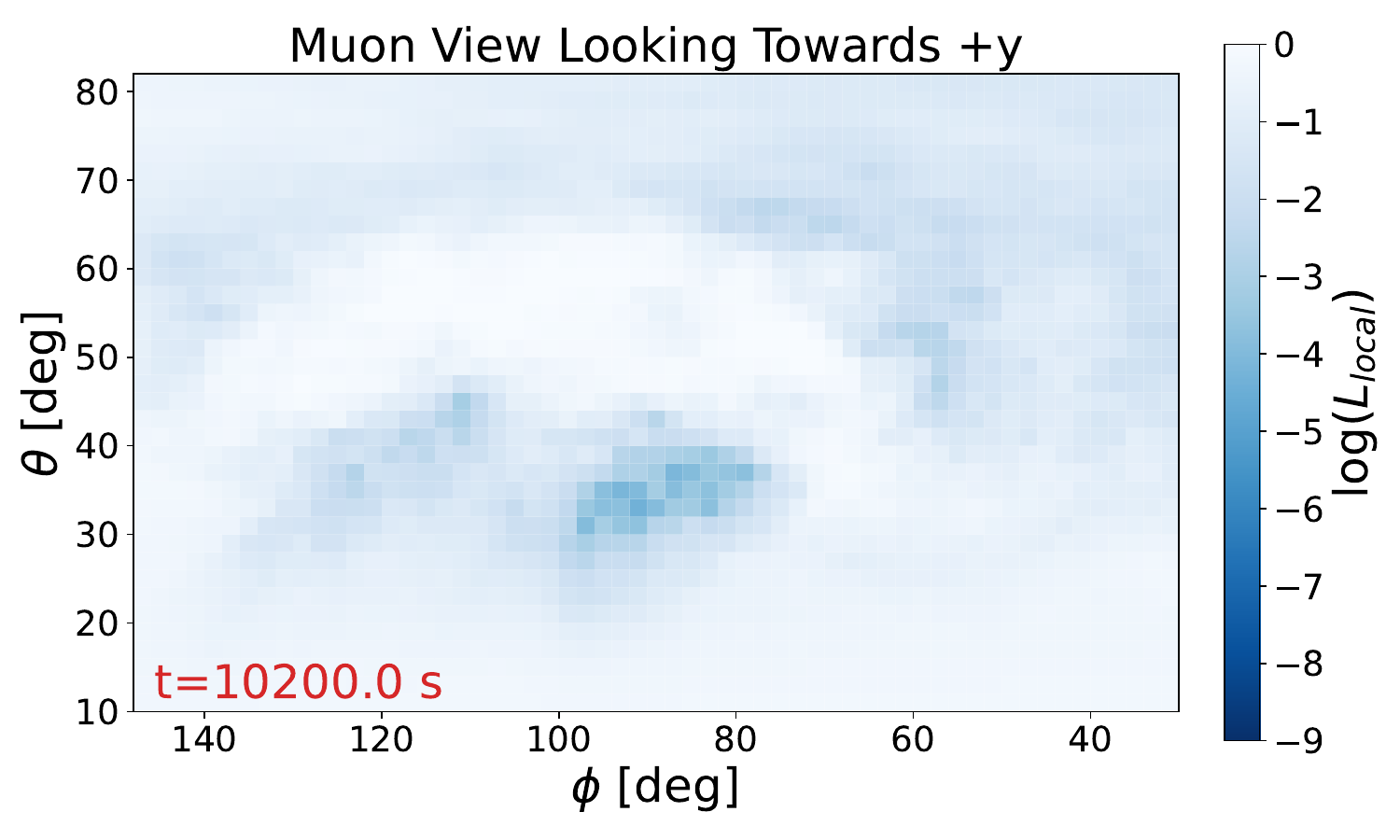}
    \caption{Expectations of the effect of a supercell thunderstorm on the atmospheric muon flux, excluding the effect of hydrometeors. Each image is obtained from integrating the muon flux over the preceding hour. In comparison with figure \ref{fig:muonflux}, the muon flux perturbation is slightly more extreme.}
    \label{fig:muflux_dryair}
\end{figure*}

Hydrometeors (rain and hail) are present in many tornadic supercells, and can have a notable effect on the density field. Water vapor makes the air density slightly lower, while solid and liquid hydrometeors make the air slightly heavier. The storm simulation presented in this paper includes the simulation of hydrometeors in the calculation of the density field, and consequently, also includes their effect on muon propagation through the storm.  To explore the effect of hydrometeors on the muographic density measurement technique presented in this paper, the muon flux calculation was repeated using the ``density of dry air" (the density excluding any hydrometeors) in place of the ``total air density" (including hydrometeors). This allows us to compare the effect on the muon flux excluding hydrometeors (the ``density of dry air" calculation), with the full effect including hydrometeors (the ``total air density" calculation) shown in previous sections of this paper. A reproduction of figure \ref{fig:muonflux} using only the density of dry air to compute the expected muon flux can be seen in figure\ref{fig:muflux_dryair}. 

The effect of hydrometeors on the muon flux calculation is present, but small. In comparison to figure \ref{fig:muonflux}, figure \ref{fig:muflux_dryair} displays a slightly larger muon flux excess from the direction of the tornado. This is expected, as the presence of hydrometeors increases the total matter density in this region, offsetting some of the effect of the lowered dry air density. The net effect is that hydrometeors reduce the efficacy of this technique, though as can be seen in figure \ref{fig:muflux_v_t} the effect of tornado formation on the atmospheric muon flux can still be seen even when using accounting for the presence of hydrometeors. 

\subsection{Other Weather Phenomena}
In the more general case, atmospheric phenomena can be characterized by their physical size and the scale of their density perturbations. By simulating various atmospheric phenomena described by these two variables, we can estimate what types of atmospheric phenomena are candidates for remote density measurements using atmospheric muons. 

\begin{figure*}
    \centering
    \includegraphics[width=0.8\textwidth]{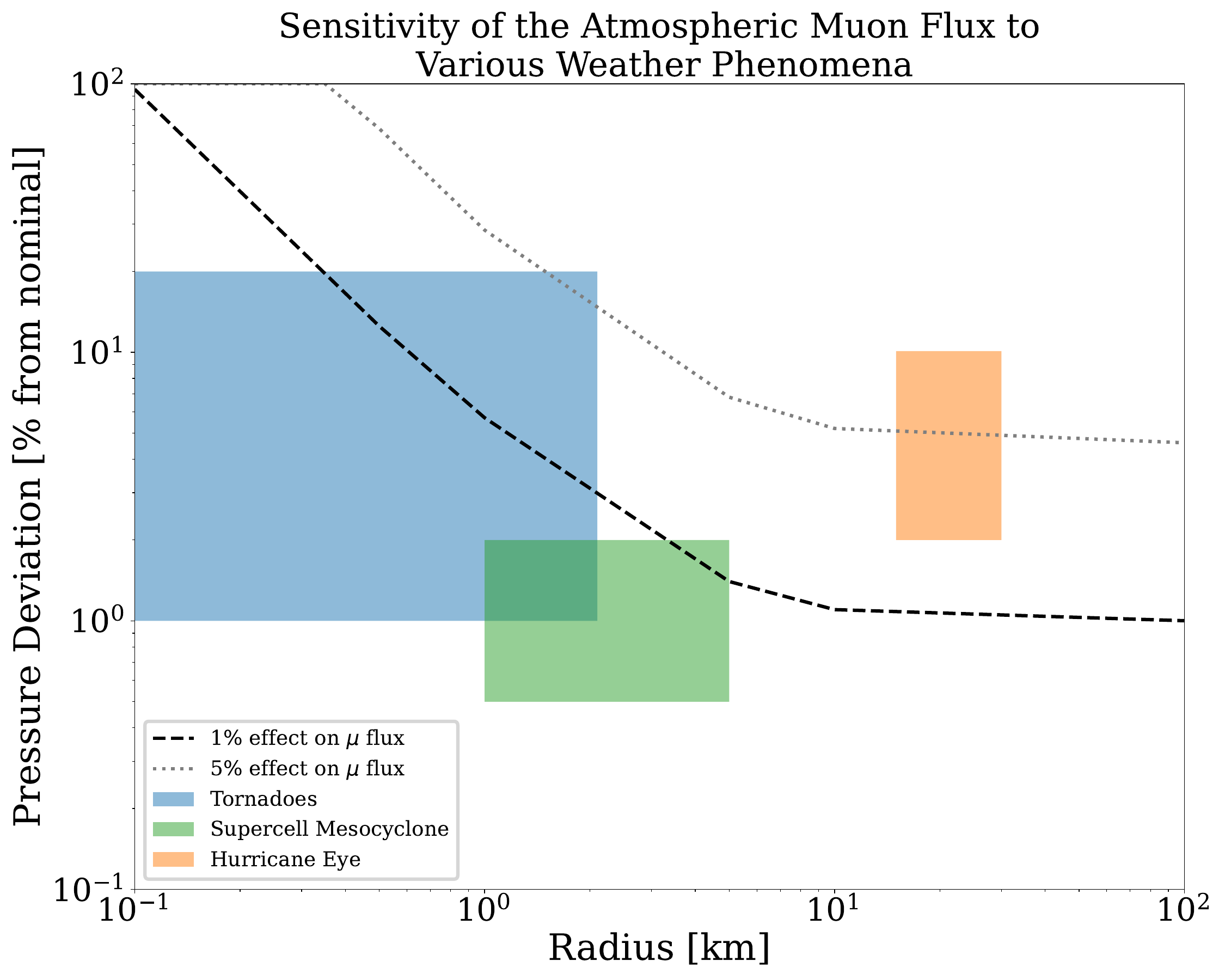}
    \caption{Various severe weather phenomena plotted in the space of size (x-axis), and magnitude of associated density drop (y-axis)~\cite{hurricanep,hurricaner,supercellevolution,tornador,hurdat}. Highlighted regions for tornadoes, thunderstorms, and hurricanes are approximate, as these values have not been exhaustively cataloged for the majority of storms. For the purposes of this plot, supercell thunderstorms are estimated to have density deviations comparable to that of tornadoes, though this has yet to be measured. Curves corresponding to regions of a 1\% and 5\% effect on the atmospheric muon flux are shown, where points above these curves are expected to produce a larger deviation in the muon flux. }
    \label{fig:sensplot}
\end{figure*}

A simplified simulation of the density field is used to achieve this. Different atmospheric phenomena are simulated as 10-kilometer tall cylinders of negative density perturbation, with variable horizontal radius. The radius and average density perturbation are varied, and for each case the muon flux residual integrated over the full field of view is recorded. This can be used to estimate how large of an effect various atmospheric phenomena might have on the muon flux. These curves, as well as highlighted regions corresponding to different atmospheric phenomena, can be seen in figure \ref{fig:sensplot}. These results seem consistent with previous studies~\cite{typhoons,muthunderstorms2}, which have observed perturbations in the muon flux in the vicinity of typhoons (hurricanes) and non-tornadic thunderstorms.

\section{Conclusion}
Here we have presented simulations of the atmospheric muon flux in the presence of a simulated supercell thunderstorm with an active tornado. We find that such a storm would introduce directional variations in the atmospheric muon flux of approximately 1\%. If a storm can be observed over the course of an hour, a 1000 m$^2$ muon detector would likely be capable of measuring variations in the local density field over approximately 1 kilometer length scales. If a prior model for the density field associated with the tornado itself can be developed, this technique could additionally be used to identify the presence and location of an active tornado. 

The 1000 m$^2$ muon detector considered in this paper is smaller than several existing cosmic ray telescopes. Detectors covering an even smaller area would likely be able to do similar science, and would have the advantage of portability, though will have to contend with proportionally larger statistical variations. Additionally, supercell thunderstorms present a high degree of storm-to-storm variability, and while in this case the density field variations appear to be associated with tornado formation, this may not be the case for all storms.  In any case, measurements of the atmospheric muon flux in the vicinity of supercell thunderstorms could provide valuable information about the density field over a large volume that cannot be obtained using existing methods. 

\section{Acknowledgements}
The authors are grateful for helpful discussions with John Beacom, Austin Cummings, Kaeli Hughes, Peter Taylor, Justin Flaherty and Paul Martini.  Leigh Orf was partially supported by NSF grant ATM-2114757.

\clearpage
\bibliography{apssamp}

\end{document}